\documentclass[a4paper,10pt]{article}
\usepackage[utf8]{inputenc}
\usepackage{amsmath,amsthm,amssymb}
\usepackage{authblk}
\usepackage{hyperref}
\usepackage{doi}
\usepackage{cite}
\usepackage{microtype}
\usepackage{stmaryrd} \usepackage{dsfont} \usepackage{graphicx}
\usepackage{enumerate}
\usepackage{authblk} \usepackage{subfigure}
\usepackage[normalem]{ulem}
\usepackage{color}

\usepackage{xcolor}

\newcommand{\changeB}[1]{#1} \newcommand{\changeC}[1]{#1} \newcommand{\changeD}[1]{#1} 

\usepackage{chngcntr}
\usepackage{apptools}

\DeclareMathOperator{\Oh}{Oh} \DeclareMathOperator{\mum}{{\mu}m} \DeclareMathOperator{\nm}{nm} \DeclareMathOperator{\thetaphob}{\theta_{phob}}
\DeclareMathOperator{\thetaphil}{\theta_{phil}}

\newcommand{\ddt}[1]{\frac{d #1}{dt}}

\newcommand{\DDT}[1]{\frac{D #1}{Dt}}

\newcommand{\jump}[1]{\left\llbracket #1 \right\rrbracket}

\newcommand{\divergence}[1]{\nabla \cdot #1}
\DeclareMathOperator{\divsigma}{div_\Sigma}

\newcommand{\transpose}{{\sf T}}

    \newcommand{\visc}{\eta}

\newcommand{\nsigma}{n_\Sigma}

\newcommand{\ndomega}{n_{\partial\Omega}}

  \newcommand{\normalspeed}{V_\Sigma}  

\usepackage[a4paper,left=2.5cm,right=2.5cm,top=3cm,bottom=3cm]{geometry}

\usepackage[english]{babel}
\addto\captionsenglish{}

\date{}

\usepackage[font=small,labelfont=bf]{caption}

\begin{document}

\title{Breakup Dynamics of Capillary Bridges on Hydrophobic Stripes}
\author[1]{Maximilian Hartmann}
\author[2]{Mathis Fricke\thanks{Corresponding auhor.}}
\author[1]{Lukas Weimar}
\author[2]{Dirk Gründing}
\author[2]{Tomislav Mari\'{c}}
\author[2]{Dieter Bothe}
\author[1]{Steffen Hardt}

\affil[1]{\footnotesize Nano- and Microfluidics Group, TU Darmstadt, Alarich-Weiss-Straße 10, 64287 Darmstadt, Germany}
\affil[2]{\footnotesize Mathematical Modeling and Analysis Group, TU Darmstadt, Alarich-Weiss-Straße 10, 64287 Darmstadt, Germany}

\maketitle

\begin{abstract}
The breakup dynamics of a capillary bridge on a hydrophobic stripe between two hydrophilic stripes is studied experimentally and numerically using direct numerical simulations.
The capillary bridge is formed from an evaporating water droplet wetting three neighboring stripes of a chemically patterned surface.
By considering the breakup process in \changeC{a phase space representation}, the breakup dynamics can be evaluated without the uncertainty in determining the precise breakup time.
The simulations are based on the Volume-of-Fluid (VOF) method implemented in Free Surface 3D (FS3D).
In order to construct physically realistic initial data for the VOF simulation, Surface Evolver is employed to calculate an initial configuration consistent with experiments.
Numerical instabilities at the contact line are reduced by a novel discretization of the Navier-slip boundary condition on staggered grids.
The breakup of the capillary bridge cannot be characterized by a unique scaling relationship. Instead, at different stages of the breakup process different scaling exponents apply, and the structure of the bridge undergoes a qualitative change. In the final stage of breakup, the capillary bridge forms a liquid thread that breaks up consistently with the Rayleigh-Plateau instability. \end{abstract}

This preprint was accepted for publication in the \emph{International Journal of Multiphase Flow}. When citing this work, please refer to the journal article: \textbf{DOI:} \href{https://doi.org/10.1016/j.ijmultiphaseflow.2021.103582}{10.1016/j.ijmultiphaseflow.2021.103582}.\newline
\newline
\textbf{Keywords:} Structured surface, Capillary bridge, Volume-of-Fluid, Breakup dynamics,\\ Rayleigh-Plateau instability
\let\thefootnote\relax\footnotetext{\textbf{E-Mail addresses:} \href{mailto:hartmann@nmf.tu-darmstadt.de}{hartmann@nmf.tu-darmstadt.de} (M.\ Hartmann),  \href{mailto:fricke@mma.tu-darmstadt.de}{fricke@mma.tu-darmstadt.de} (M.\ Fricke), \href{mailto:lukas.weimar@stud.tu-darmstadt.de}{lukas.weimar@stud.tu-darmstadt.de} (L.\ Weimar), \href{mailto:gruending@mma.tu-darmstadt.de}{gruending@mma.tu-darmstadt.de} (D.\ Gründing), \href{mailto:maric@mma.tu-darmstadt.de}{maric@mma.tu-darmstadt.de} (T.\ Mari\'{c}), \href{mailto:bothe@mma.tu-darmstadt.de}{bothe@mma.tu-darmstadt.de} (D.\ Bothe), \href{mailto:hardt@nmf.tu-darmstadt.de}{hardt@nmf.tu-darmstadt.de} (S.\ Hardt)}

\section{Introduction}
Wetting of patterned surfaces is omni-present in nature.
The Lotus effect~\cite{Barthlott1997} or the fog harvesting of the \textit{Stenocara} desert beetle in the Namib Desert~\cite{Parker2001} are only two examples.
Also when it comes to engineering applications like ink-jet printing~\cite{Wang2004} or water harvesting~\cite{Zhang2015a}, the understanding of wetting behavior on (chemically) patterned surfaces is of crucial importance.

In the present study, the focus lies on chemically patterned striped surfaces that are wetted by water droplets with radii of the order of magnitude of the stripe width.
For such kind of systems, static wetting behavior has been studied by using energy minimization techniques~\cite{David2012,Matsui2012,He2018}, lattice Boltzmann simulations \cite{Jansen2013,Jansen2016} and experiments~\cite{Bliznyuk2009}.
Also inertia-driven spreading~\cite{Jansen2016} and the splitting~\cite{Leopoldes2003,Song2015,Zou2018} of impinging droplets on such kind of surfaces have been investigated in some detail.
When it comes to evaporation, simulations using the phase field~\cite{Wu2019} method have been performed more recently.
Hartmann and Hardt~\cite{Hartmann2019} showed that an evaporating droplet wetting two hydrophilic stripes, with a hydrophobic one in between them, is stable as long as the pressure inside the liquid forming the capillary bridge above the hydrophobic stripe can be balanced in the liquid above the two hydrophilic stripes.
This is the case until a certain width of the capillary bridge is reached.
When the width of the bridge decreases further due to evaporation, it breaks up.
While in the latter publication the focus is on the statics of an evaporating droplet, the subject of the present article is the dynamics of the breakup process itself.

Qualitatively, this process shows similarities to the collapse of a soap film between two circular rings, as it was investigated by Chen and Steen~\cite{Chen1997} using numerical calculations.
The authors found two breakup regimes, \changeC{a $d \sim \tau^{2/3}$-, and a $d \sim \tau^{2/5}$-regime (with $d$ the minimum bridge width and $\tau$ the time before breakup)}, which are both dominated by the balance between capillary and inertial forces.
The change in regime is due to a geometric transition during breakup.
The \changeC{$d \sim \tau^{2/3}$-}scaling is a classical result for the evolution of the minimum width $d$ of a ``free liquid bridge'' (i.e.\ without contact to a substrate) obtained from dimensional analysis~\cite{Keller1983}
\begin{align}
\label{eqn:inviscid_breakup_power_law}
d(\tau)  = C \left(\frac{\sigma \tau^2}{\rho}\right)^{1/3},
\end{align}
where $\sigma$ is the surface tension, $\rho$ is the fluid density and $\tau = t_0 -t$ is the time $t$ before the bridge breaks up at $t_0$.
The prefactor $C$ has been believed to be a universal constant with a value close to $1.4$ for quite some time (see, e.g., Eggers and Fontelos \cite{Eggers2015}). Assuming the prefactor to be universal with $C=0.9 \pm 0.01$, Hauner et al.\ \cite{Hauner2017} concluded that water exhibits a dynamic surface tension, characterizing freshly created surface sections, of $\sigma \approx 90 \, \text{mN}/\text{m}$, which is significantly larger than the equilibrium value of $72 \, \text{mN}/\text{m}$. However, more recent studies show that $C$ is indeed not universal. Instead, the experimentally observable value of the prefactor may depend on the fluid and system parameters \cite{Deblais2018}.\\
\\
According to Li \& Sprittles \cite{Li2016}, the appropriate dimensionless number characterizing the breakup of a free capillary bridge of a liquid is the Ohnesorge number
\begin{align}
\Oh = \frac{\visc}{\sqrt{\rho \sigma R}} \, ,
\end{align}
which is the square root of the ratio of the viscous length scale $\visc^2/(\rho \sigma)$ (with $\visc$ being the dynamic viscosity) and a characteristic length scale $R$. We choose \changeC{the width of the hydrophobic stripe $250 \mum \leq w_{\text{phob}} \leq 750 \mum$} as the characteristic length scale \changeC{i.e., $R = w_\mathrm{phob}$. With this definition, the Ohnesorge number for water in air for the relevant hydrophobic stripe widths lies within the range}
\begin{align}
\changeC{4.3 \cdot 10^{-3} \le \Oh \le 7.5 \cdot 10^{-3}}.
\label{eq:OhOurCase}
\end{align}

Clearly, the limiting cases $\Oh \rightarrow 0$ and $\Oh \rightarrow \infty$ characterize the inviscid and viscous breakup regime, respectively.
The inviscid breakup regime is already mentioned above for the soap film.
Within the viscous regime, viscous forces play the dominant role and $d$ scales linearly with time $\tau$, $d \sim K \tau$, with $K$ being a constant.
In the same publication \cite{Li2016}, a phase diagram is presented, in which, besides the above-mentioned two regimes, also a third regime, the viscous inertial regime for intermediate values of $\Oh$, can be identified in the parameter space that is spanned by $d$ and $\Oh$.
Here, as well as in the viscous regime, $d \sim K \tau$ is valid but with a different constant $K$.
Li and Sprittles \cite{Li2016} performed numerical simulations for a free capillary bridge with a similar value of $\Oh$ compared to equation~\eqref{eq:OhOurCase} and observed dynamic transitions into the inviscid regime and from the inviscid into the viscous regime as the minimum diameter of the bridge decreases approaching breakup.\\
\\
Note that Chen and Steen~\cite{Chen1997} as well as Li and Sprittles~\cite{Li2016} deal with a free capillary bridge without substrate contact.
This is not the case in the present investigation, where the capillary bridge is in contact with the substrate.
For liquids in contact with a substrate, Bostwick and Steen~\cite{Bostwick2015} provided a review, where the focus lies on the stability of constrained capillary surfaces in general.
The same authors theoretically investigated the instability of static rivulets~\cite{Bostwick2018} and considered varicose (symmetric) and sinuous (anti-symmetric) modes for pinned and free contact lines.
For symmetric modes, they confirmed the results by Davis~\cite{Davis1980}:
For all contact angles, breakup can happen in the case of a free, i.e. unpinned, contact line, while for pinned contact lines, contact angles must be greater than $90^\circ$.
Other studies exist for unpinned \cite{Dziedzic2019,Brinkmann2005b} and pinned \cite{Speth2009} contact lines, as well.
In the case of anti-symmetric modes, a static rivulet is always stable in the case of a pinned contact line and unstable only for contact angles larger than $90^\circ$.\\
\\
The above-mentioned articles deal with the stability of rivulets.
To the best knowledge of the authors, no investigations of the breakup and dynamics of capillary bridges in contact with surfaces have been published so far. 
The aim of this study is to examine the corresponding breakup dynamics and to compare it to a ``free'' liquid bridge. A collection of the reported data is published in an
open research data repository \cite{OnlineRepo}. 
\section{Experimental Methods}
\label{sec:expDetails}
Substrates are prepared by creating a stripe pattern of a positive photo resist (AZ 9260, MicroChemicals GmbH, Germany) on a borofloat33 glass wafer (Siegert Wafer GmbH, Germany) using standard photolithography steps.
This allows to produce covered and uncovered glass areas.
After oxygen plasma treatment, these wafers are silanized in a low-pressure chemical vapor deposition process, for which 1H,1H,2H,2H-Perfluorodecyltrichlorosilane (PFDTS, CAS: 78560-44-8, abcr GmbH, Germany) is used.
In the following step, the photo resist is rinsed off using acetone and isopropanol.
Subsequently, the substrate is dried in a nitrogen stream.
This leads to a pattern of hydrophilic and hydrophobic stripes.
The substrates are then stored until experiments are performed.
Different hydrophilic stripe widths $w_\mathrm{phil}$ and different ratios $\alpha$ of hydrophobic and hydrophilic stripe widths are used, where
\[ \alpha = w_\mathrm{phob}/w_\mathrm{phil}.\]

Experiments are performed by placing de-ionized water droplets \changeC{(Milli-Q device; specific resistance 18.2 M$\Omega \cdot$cm at 25~$^\circ$C)} with a pipette onto the stripe pattern.
The volume is chosen in a way that two hydrophilic stripes with one hydrophobic stripe in between them are wetted.
The water then evaporates in the lab environment (temperature $\sim$25~$^\circ$C) until a critical width of the bridge on the hydrophobic stripe is reached. The critical width marks the transition to a configuration beyond which no stationary wetting state of the droplet exists (see Hartmann and Hardt~\cite{Hartmann2019} for a discussion of the stability of the liquid bridge).
We did not control the lab humidity since the final breakup process occurs at a small time scale which we expect to be of the order of the capillary time scale.
It can be shown that the capillary time scale is about 5 orders of magnitude smaller than the evaporation time scale that therefore can be neglected in the present case (see Appendix~\ref{section:timescale_estimation} for more details).

The final time span ($\sim$ 0.13 s) before breakup, i.e.\ the time span in which the capillary bridge on the hydrophobic stripe decays, is recorded with a high-speed camera (Photron FASTCAM SA-1.1) in top view mode.
In order to obtain the values necessary for fixing the initial condition for the numerical simulations, e.g. the wetted length and the contact angle on the hydrophilic stripe in the moment of breakup, a second high-speed camera (Photron FASTCAM SA-X2) is used, which records the droplet from one side.
The cameras are synchronized via dedicated software (Photron FASTCAM Viewer) and triggered externally.
A frame rate of 75,000 frames per second (fps) is used.
Each camera is connected to a macro objective (Navitar-12X).
Illumination is performed via backlight in side view and co-axially in top view using two cold-light sources (VOLPI intra LED 5).
A silicon wafer that acts as a mirror is placed below the glass substrate in order to achieve good illumination conditions and record the breakup process at a sufficient frame rate and magnification.
We will also report results on the very last stages of bridge breakup.
In these cases, the capillary bridge is recorded in bottom view mode with the Photron camera attached to a microscopy body.
A more detailed description of this experimental setup, used materials and substrate preparation steps can be found in a previously published article~\cite{Hartmann2019}.
Contact angles on the silanized glass wafers are measured using the evaporation method, in which the receding contact angle can be measured in the constant-contact angle mode during evaporation~\cite{Park2012}. 
\section{Numerical Methods}
\subsection{Stationary States from Energy Minimization} \label{sec:minimal_surfaces}
To be able to simulate the process of droplet breakup, a precise geometric description of the liquid distribution at the beginning of the simulation is indispensable.
Hence, the interface geometry has to be extracted from the experimental observations and transformed into a phase fraction field that can be processed by Free Surface 3D (FS3D).
For this purpose, a triangulated surface is produced using Surface Evolver \cite{Brakke1992}, a tool for calculating minimal surfaces by minimizing the free energy functional \changeC{that, in the present case, can be written as}
\begin{equation}
\changeC{\mathcal{F}= \sigma_\mathrm{lg} \cdot |A_\mathrm{lg}| + \sum_{i \in \{ \mathrm{phil, phob} \}} (\sigma_{\mathrm{sl,}i} - \sigma_{\mathrm{sg,}i})\cdot|A_{\mathrm{sl,}i}|}
\label{eq:energyChangeSE}
\end{equation}
subject to a prescribed volume \changeC{$V$}, contact angle \changeC{$\Theta$} and wetted length on the hydrophilic stripe \changeC{$l_\mathrm{phil}$}.
In the present case, the interfacial tension between liquid and gas \changeC{$\sigma_\mathrm{lg}$} is constant, while the specific interfacial areas $|A_\mathrm{j,i}|$ can change to minimize the free energy $\mathcal{F}$.
The subscripts sl, \changeC{sg, and lg denote solid-liquid, solid-gas and gas-liquid, respectively.
Since the droplet wets alternating hydrophilic and hydrophobic (abbreviated with phil and phob) stripes, both must be taken into account for calculating $\mathcal{F}$.
Note that $\sigma_{\mathrm{sl,}i} - \sigma_{\mathrm{sg,}i}$ can be substituted by Young's equation, which is useful in the present case, because the contact angles on $i \in \{ \mathrm{phil, phob} \}$ can be determined experimentally.}\\

\begin{figure}[t]
 \centering
  \includegraphics[width=0.5\textwidth]{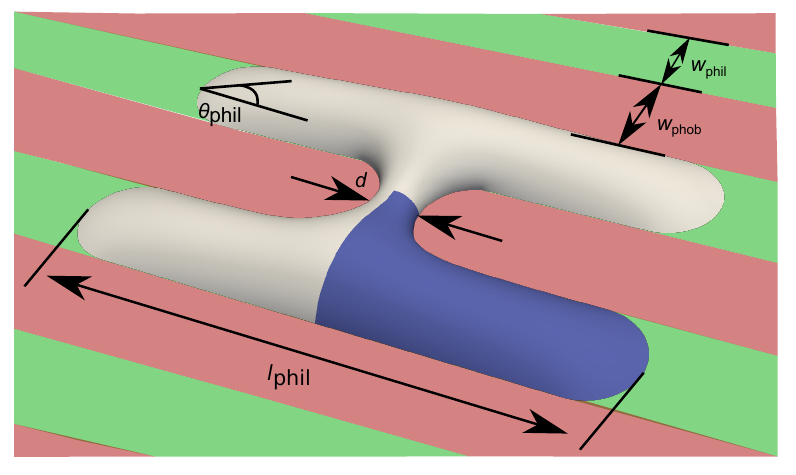}
 \caption{Initial geometry for the numerical simulations obtained from Surface Evolver (blue). Due to symmetry, only 1/4th of the droplet is calculated. Red areas denote hydrophobic, green areas hydrophilic stripes. The wetted length $l_\mathrm{phil}$, the contact angle on the hydrophilic stripe $\Theta_\mathrm{phil}$, the hydrophilic stripe width $w_\mathrm{phil}$ and $\alpha$, the ratio between the hydrophobic and the hydrophilic stripe width, are held constant during the calculation. Used values can be found in Table~\ref{tab:initGeo}.}
 \label{fig:surfaceEvolver_initGeo}
\end{figure}

As input data, the wetted length of the hydrophilic stripe $l_\mathrm{phil}$ as well as the contact angle $\Theta_\mathrm{phil}$ on it, as observed in the experiments, are used.
These values are summarized in Table~\ref{tab:initGeo}.
\begin{table}[b]
\centering
\begin{tabular}{lcc}
\hline
$\alpha~(-)$ & $\Theta_\mathrm{phil}~(^\circ)$ & $l_\mathrm{phil}~(\mum)$\\
\hline
0.5 & 21 & 1966\\
1   & 28 & 2638\\
1.5 & 31 & 2936\\
\hline
\end{tabular}
\caption{Contact angle and wetted length on the hydrophilic stripe as observed in the experiments and used in the Surface Evolver calculations.}
\label{tab:initGeo}
\end{table}
Furthermore, the contact angle on the hydrophobic stripe $\Theta_\mathrm{phob}$ is assumed to be the receding contact angle on a silanized glass wafer.
The corresponding experimental value is 102$^\circ$.
All of these three parameters are kept constant during the whole minimization procedure.
Note that at the boundary between the hydrophilic and the hydrophobic stripe, the contact angle may lie between $\Theta_\mathrm{phil}$ and $\Theta_\mathrm{phob}$.
Due to symmetry considerations and to reduce calculation time, only a quarter of a droplet is calculated\changeC{, as it is indicated with blue color in Fig.~\ref{fig:surfaceEvolver_initGeo}. 
For that purpose, two mirror planes are considered.
Both are oriented perpendicular to each other and to the surface itself.
While one mirror plane is located parallel to the stripes in the middle of the hydrophobic one, the other plane is oriented perpendicular to the stripes, cutting the hydrophobic bridge in the middle along its axis.}
At these mirror planes, the contact angle is set to 90$^\circ$. \changeC{This means that the liquid–gas interface meets the symmetry plane at an angle of 90$^\circ$, which is a symmetry boundary condition.
}\\

\changeC{
The goal of the Surface Evolver calculations is to find the equilibrium solution of the capillary surface that obeys the above described boundary conditions (i.e. wets both the hydrophilic and the hydrophobic stripes) and belongs to the smallest volume of liquid, which corresponds to the minimum stable liquid bridge configuration.
For that purpose an algorithm is implemented as described in the following:
}
\begin{enumerate}
\item{Start calculation with a certain volume \changeB{$V$. If this is the initial step of the droplet evolution, the volume is $V = V_0$.}}
\item{Evolve surface and calculate
\changeC{
\begin{equation}
c_{v,n} = \frac{s_n}{\overline{\mathcal{F}}} = \frac{\frac{1}{\sqrt{n(n-1)}}\sqrt{\sum_i^{i-n} (\overline{\mathcal{F}} - \mathcal{F}_i)^2}}{\frac{1}{n}\sum_i^{i-n}\mathcal{F}_i},
\label{eq:coeffOfVari}
\end{equation}
the coefficient of variation of the energy of the $n$ last iterations.
$c_{v,n}$ is the ratio between the standard deviation $s_n$ and the mean value of the free energy $\overline{\mathcal{F}}$ at iteration $i$.
Note that in the present work $n=7$.}}
\item{\changeC{Calculate bridge width $d$.}}
\item{Check for convergence. Convergence is reached when $c_{v,7} < 10^{-7}$. If the calculation has converged, \changeB{save $V$ as $V_\mathrm{stable}$, and} decrease volume \changeB{by $\Delta V = V_0 / 2$ if convergence has been reached for the first time, or by $\Delta V = \Delta V / 2$ if convergence has been reached at least once, so that $V = V - \Delta V$. Then} check if the criterion of step \changeC{6} is fulfilled. If not, continue with step \changeC{5}.}
\item{Check for breakup. The capillary bridge is considered to be broken up if $d < 0.1\cdot w_\mathrm{phil}$. This value is well below the critical bridge width according to Hartmann and Hardt~\cite{Hartmann2019}}. If this criterion is fulfilled, increase volume \changeB{by $\Delta V = \Delta V / 2$ so that $V = V + \Delta V$}, and check if the criterion of step \changeC{6} is fulfilled. If not, repeat steps 1 - \changeC{5}.
\item{Check if \changeB{$\Delta V$} is smaller than 0.5\,\% of the last stable volume \changeB{$V_\mathrm{stable}$}. If yes, stop the iteration.}
\end{enumerate}
During the evolution of the surface, the mesh is successively refined until the \changeC{maximum length of each cell edge} is smaller than $1/100\cdot(w_\mathrm{phil}+w_\mathrm{phob})$. Convergence can only be achieved if this is true.
\changeB{The overall procedure of volume variation, as described above, follows the principle of nested intervals and is similar to a binary search algorithm.}
\changeD{From the stopping criterion $\Delta V = 0.005 \cdot V_\mathrm{stable}$ it is expected that the last found critical volume is close enough to the real value.
Note that it is impossible to calculate arbitrarily close to the real critical volume.}
The result of the algorithm is the triangulated surface mesh with the smallest stable volume.
This geometry is exported as a triangulated surface mesh in STL (Standard Triangulation Language) format and subsequently transformed into a volume-fraction field for FS3D (see Section~\ref{subsection:vof}).
Note that Surface Evolver is a tool that is usually used to calculate stable minimal surfaces. 
In the present case, it is used to calculate an equilibrium shape close to the \changeC{onset of instability}.
Moreover, from experiments it is known that before the capillary bridge becomes unstable (during evaporation), the contact angle on the hydrophobic stripe changes within the range that is given by the advancing and receding contact angle.
In Surface Evolver, the contact angles are set to fixed values, i.e. contact angle hysteresis is not accounted for, although it is observed in the experiments.
Therefore, the initial configuration differs from the one observed in the experiments (see Fig.~\ref{fig:initCondition} in Appendix~\ref{section:initial_condition}).

\subsection{Continuum Mechanical Model}
\begin{figure}[ht]
 \centering
 \includegraphics[width=8cm]{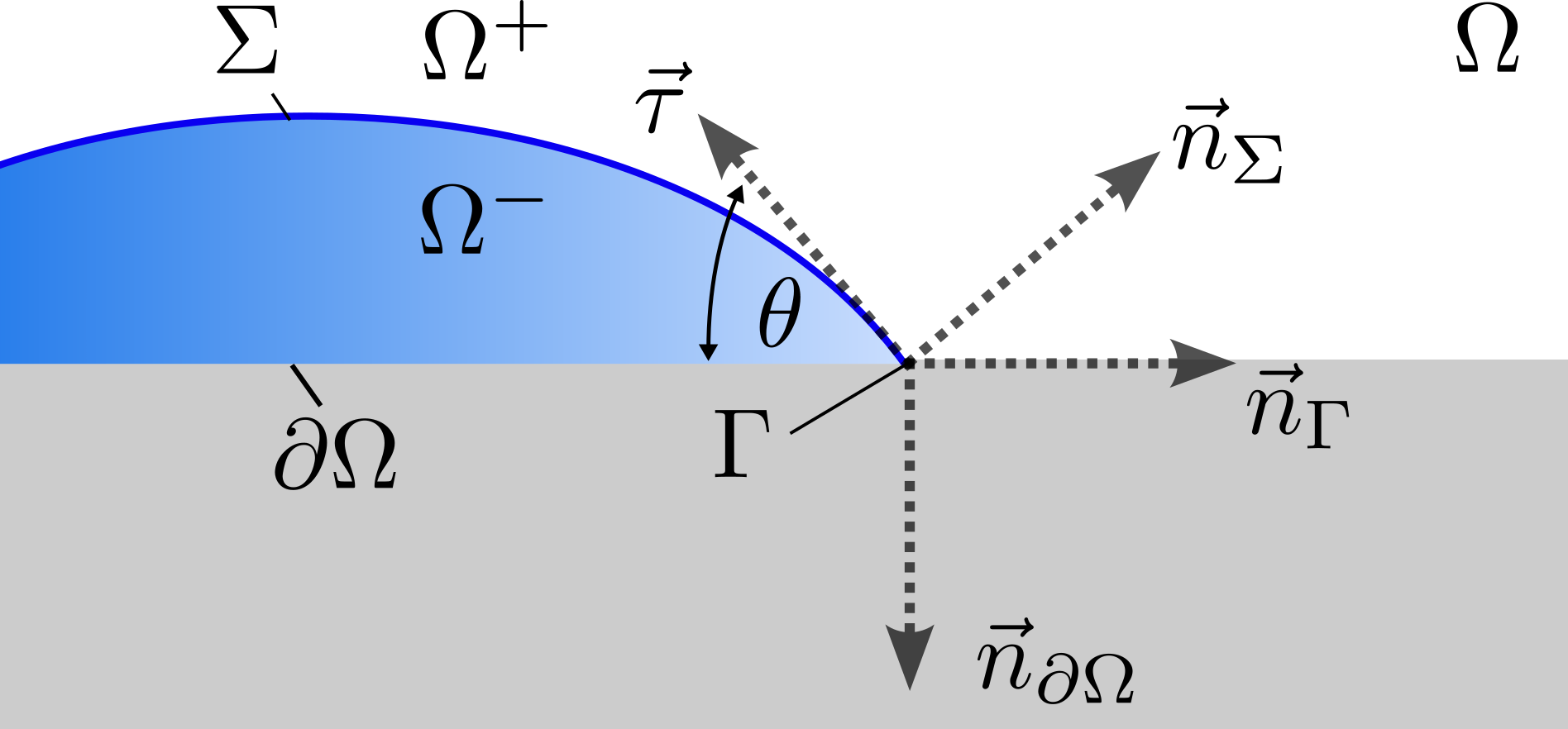}
 \caption{Illustration of employed mathematical notations.}
\end{figure}
The continuum mechanical model is based on the incompressible two-phase Navier Stokes equations in the sharp interface formulation. The conservation equations for momentum and mass in the bulk phases read
\begin{align}
\label{eqn:bulk_equations}
\rho \DDT{\vec{v}} - \divergence{S} + \nabla p = \rho \vec{g}, \ \nabla \cdot \vec{v} = 0 \ \quad &\text{in} \ \Omega\setminus\Sigma(t),
\end{align}
where $p$ is the pressure and $S = \visc (\nabla \vec{v} + {\nabla \vec{v}}^\transpose)$ is the viscous stress tensor. Note that $\vec{g}$ is the gravitational acceleration and the Lagrangian time-derivative is defined as
\[ \DDT{} = \partial_t + \vec{v} \cdot \nabla \,. \]
To simplify the model, it is assumed that no mass is transferred across the gas-liquid interface. This can be expected to be a good approximation on the timescale of the breakup process, which is much smaller than the evaporation timescale (see estimation in Appendix~\ref{section:timescale_estimation}). Together with the assumption of no slip at the gas-liquid interface, we obtain continuity of the velocity field $v$, i.e.\,
\begin{align}
\label{eqn:velocity_jump}
\jump{\vec{v}} = 0 \quad \text{on} \ \Sigma(t),
\end{align}
where $\jump{\psi}(t,x) = \lim_{h \rightarrow 0^+} \psi(t,x+h\nsigma) - \psi(t,x-h\nsigma)$ denotes the jump of a given (discontinuous) quantity across the interface. The interfacial transmission condition for momentum in case of constant surface tension reads
\begin{align}
\label{eqn:momentum_jump}
\jump{p \mathds{1} - S} \, \nsigma = \sigma \kappa \nsigma \quad &\text{on} \ \Sigma(t).
\end{align}
The right-hand side is the surface tension force with $\sigma > 0$ the surface tension and $\kappa = - \divsigma \nsigma$ the mean curvature. The motion of the interface is coupled to the bulk flow through the kinematic boundary condition
\begin{align*}
\normalspeed = \vec{v} \cdot \nsigma \quad &\text{on} \ \Sigma(t),
\end{align*}
where $\normalspeed$ is the interface normal velocity. Formally, the latter condition can be reformulated as the advection equation
\begin{align}
\label{eqn:chi_transport_equation}
\partial_t \chi + \vec{v} \cdot \nabla \chi = 0,
\end{align}
where
\begin{align*}
\chi(t,x) = \begin{cases} 1 & \text{if} \quad x \in \Omega^-(t), \\ 0 & \text{if} \quad x \notin \Omega^-(t) \end{cases}
\end{align*}
is the indicator function for the liquid phase. The transport equation \eqref{eqn:chi_transport_equation} forms the basis for the Volume-of-Fluid interface capturing method (see Section~\ref{subsection:vof}). In order to regularize\footnote{\changeB{Note that the Navier slip condition with $L>0$ modifies the singularity at the moving contact line such that it becomes integrable and the viscous dissipation rate becomes finite. On the other hand, it has been shown mathematically that the solution cannot be completely regular if the slip length is finite at the contact line \cite{Fricke.2019}.}} the moving contact line singularity \cite{Huh.1971}, the Navier slip boundary condition is applied for the velocity at the solid wall, i.e.\,
\begin{align}
\vec{v} \cdot \ndomega = 0, \quad \beta \vec{v}_\parallel + (S \ndomega)_\parallel = 0 \quad &\text{on} \ \partial\Omega\setminus\Gamma(t),
\end{align}
where $L = \eta/\beta > 0$ is the so-called slip length. \changeC{For simplicity, we assume that the slip length is the same in the liquid and in the gas phase, i.e.\ $L=L^\pm$.} The wettability of the solid is modeled through the contact angle boundary condition
\begin{align}
\nsigma \cdot \ndomega = - \cos \theta \quad \text{on} \ \Gamma(t),
\end{align}
where the contact angle may be a function of position, modeling regions of different wettability. Here, for simplicity, we apply a fixed contact angle in both the hydrophilic part of the boundary $\partial\Omega_{\text{phil}}$ and the hydrophobic part of the boundary $\partial\Omega_{\text{phob}}$, i.e.\,
\begin{align*}
\nsigma \cdot \ndomega = \begin{cases} - \cos \thetaphil & \text{on} \quad \Gamma(t) \cap \partial\Omega_{\text{phil}} \\ - \cos \thetaphob & \text{on} \quad \Gamma(t) \cap \partial\Omega_{\text{phob}} \end{cases},
\end{align*}
where $0 < \thetaphil < \thetaphob < \pi$ are constants. Note that this is a simplified model, neglecting, e.g., contact angle hysteresis.

\subsection{Volume-of-Fluid Discretization}
\label{subsection:vof}

In the present study, the two-phase flow solver \emph{Free Surface 3D} (FS3D), originally developed by Rieber \changeD{and Frohn} \cite{Rieber2004,Rieber1999}, is employed to solve the incompressible two-phase Navier Stokes equations. Here we only discuss the most important aspects of the discretization and refer to the literature \cite{Fricke2021,Rieber2004,Fath2015,Fath2016,Fricke.2019b} for details.\\
\\
The two-phase Navier Stokes equations in the Continuum Surface Force (CSF) formulation \cite{Brackbill.1992} are discretized using the finite volume approach on a fixed Cartesian grid. Within the CSF formulation, the effect of surface tension is modeled as a singular source term in the Navier Stokes equations and the equations \eqref{eqn:bulk_equations} and \eqref{eqn:momentum_jump} are replaced by
\begin{align}
\rho \DDT{\vec{v}} - \divergence{S} + \nabla p &= \rho \vec{g} + \sigma \kappa \nsigma \delta_\Sigma, \quad \nabla \cdot \vec{v} = 0 \quad \text{in} \quad \Omega,
\end{align}
where $\delta_\Sigma$ denotes the surface delta distribution on $\Sigma(t)$. The phase indicator function $\chi$ is replaced by the discrete volume fraction 
\begin{align} f = \frac{1}{|V_0|} \int_{V_0} \chi \, dV \label{eq:def_volume_fraction}\end{align}
in each computational cell, which is used to track the location of the interface. Integration of \eqref{eqn:chi_transport_equation} over a control volume $V_0$ yields the transport equation for the volume fraction, i.e.\
\begin{align}
\label{eqn:f_transport_equation}
\ddt{f} = - \frac{1}{|V_0|} \int_{\partial V_0} \chi \, \vec{v} \cdot \vec{n} \, \text{dA}. 
\end{align}
Within the geometrical Volume-of-Fluid (VOF) method, the flux on the right-hand-side of \eqref{eqn:f_transport_equation} is approximated by means of geometrical methods. Based on a reconstructed interface geometry, the numerical fluxes for the volume fraction are computed using an operator-splitting method \cite{Strang1968}, which decomposes the transport problem in a series of one-dimensional transport problems along the coordinate axes.  The interface is locally reconstructed as a plane in each cell (also known as piecewise linear interface calculation - PLIC) \cite{Rider1998}, where the Youngs method \cite{Youngs1984} is used to estimate the interface normal vector based on the volume fraction field. At the contact line, we employ a three-dimensional variant of the \emph{Boundary Youngs reconstruction method} \cite{Fricke.2019b}.\\
\\
Within the CSF formulation, the two-phase flow is treated as a single fluid where the density $\rho$ is volume averaged according to
\[ \rho = f \rho_l + (1-f) \rho_g.  \]
Here the subscripts $l$ and $g$ refer to the liquid and gas phases, respectively. The averaging of the viscosity in interface cells involves both arithmetic and harmonic averaging \cite{tryggvason1998computations,Focke2012}.\\
\\
The time integration is based on an explicit Euler method, where the pressure-velocity coupling is realized by Chorin's projection method which leads to an elliptic equation for the pressure to perform a projection onto the space of solenoidal velocity fields. Note that the grids for velocity and pressure are staggered to enhance the stability of the method \cite{Harlow1965}.\\
\\
The surface tension force is discretized with the \emph{balanced} CSF method introduced by Popinet \cite{Popinet2009}. A height-function representation of the interface is constructed in order to approximate the mean curvature. It has been demonstrated that this method is able to significantly reduce spurious currents at the interface away from the boundary. Following the approach by Afkhami and Bussmann \cite{Afkhami2008,Afkhami2009}, the height function is also used to indirectly enforce the contact angle boundary condition. The idea is to linearly extrapolate the height function at the contact line into a ghost cell layer, where the slope of the extrapolated interface is determined by the prescribed contact angle. As a result, the approximated value of the mean curvature is altered, leading to a ``numerical force'' that drives the interface towards the desired contact angle. A drawback of this method is that it may create spurious currents at the contact line.\\
\\
In order to ensure the stability of the numerical method, the time step is chosen according to the stability criterion
\[ \Delta t = \min \{ (\Delta t)_\sigma, (\Delta t)_\visc, (\Delta t)_{\boldsymbol{v}} \} \]
with the timescales (see Tryggvason et al. \cite{Tryggvason2011} for a similar criterion) given by
\[ (\Delta t)_\sigma = \sqrt{\frac{(\rho_l+\rho_g)(\Delta x)^3}{4\pi\sigma}}, \quad (\Delta t)_\visc = \frac{\rho_l (\Delta x)^2}{6 \visc_l}, \quad (\Delta t)_{\boldsymbol{v}} = \frac{\Delta x}{\lVert \vec{v} \rVert}_\infty.  \]

\paragraph{No-slip and Navier Slip boundary conditions:} To allow for a motion of the contact line, one can either make use of the \emph{numerical slip} inherent to the method or prescribe the (staggered) Navier slip condition. The numerical slip is a property of the advection algorithm since the latter uses face-centered values to transport the volume fraction field. In fact, the velocity boundary condition is indirectly enforced within the finite volume method using the concept of ``ghost cells''. 

\begin{figure}[ht]
 \centering
 \subfigure[No-slip (``Numerical slip'').]{\includegraphics[height=3cm]{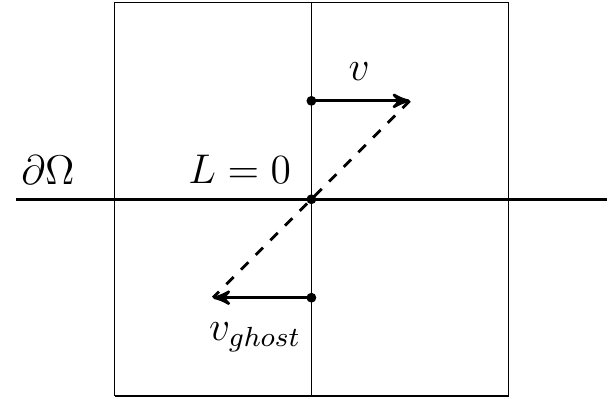}\label{fig:no-slip}}
 \subfigure[Standard Navier slip.]{\includegraphics[height=3cm]{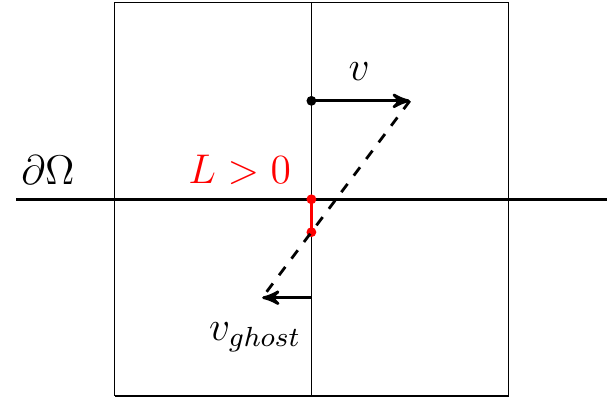}\label{fig:navier-slip}}
 \subfigure[\emph{Staggered} Navier slip.]{\includegraphics[height=3cm]{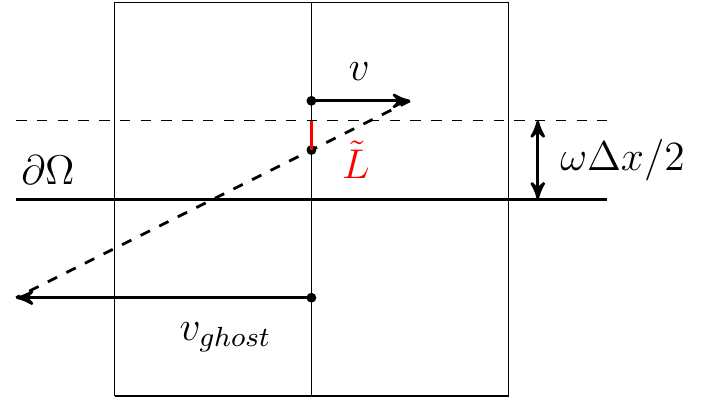}\label{fig:staggered-slip}}
 \caption{Ghost-cell based numerical realization of the no-slip, Navier slip and \emph{staggered} Navier slip boundary conditions in FS3D (for (a) and (b) see also Gründing et al.\ \cite{Gruending.2020}).}
 \label{fig:slip_in_fs3d}
\end{figure}

\begin{enumerate}[(i)]
 \item In the classical approach, the no-slip boundary condition is enforced by extrapolating the tangential velocity field into the layer of ghost cells next to the physical boundary, such that the \changeD{tangential} velocity interpolates to zero exactly at the boundary (see Fig.~\ref{fig:no-slip}), i.e.\
 \[ v_{\text{ghost}} = - v. \]
 \changeC{The velocity in the ghost cells modifies the discrete viscous forces at the solid boundary such that the motion of the contact line is inhibited.} This approach enforces the no-slip condition only in the limit of the mesh size going to zero. Therefore, the face-centered velocity at the boundary cell layer may be non-zero leading to a motion of the contact line. However, the extent of numerical slip decreases with increasing mesh resolution, \changeD{typically} leading to a significant mesh dependence of the solution \changeD{(see, e.g, \cite{Gruending.2020})}. This numerical effect has been first described in the context of VOF methods by Renardy et al.\ \cite{Renardy.2001}.
 \item The Navier slip boundary condition is \changeD{implemented} following the same approach by setting a different velocity in the ghost-cell layer, such that the tangential velocity interpolates to zero at a fixed distance $L > 0$ from the physical boundary (see Fig.~\ref{fig:navier-slip}), i.e.\ 
 \begin{align} 
 \label{eqn:ghost_velocity_navier_slip}
 v_{\text{ghost}} = v \ \frac{2L - \Delta x}{2L + \Delta x}.
 \end{align}
 It has been demonstrated in the literature \cite{Sprittles2012,Gruending.2020} that this may lead to mesh-convergent results if the resolution of the computational mesh is well below the slip length $L$. However, the physically expected slip length is on the scale of nanometers \cite{Neto.2005}. \changeC{Since the macroscopic length scale of the problem is one the $\text{mm}$ scale, we cannot resolve the nanometer scale} with the present numerical method without massive computational costs. Moreover, note that for $L>0$ the magnitude of the ``counter velocity'' $v_{\text{ghost}}$ in the ghost cell is always less than or equal to $|v_{\text{ghost}}|$ in the case of no-slip.
\end{enumerate}

\paragraph{``Staggered'' slip:} The goal of the ``staggered slip'' boundary condition, which is introduced below, is to reduce the amount of artificial numerical slip \changeB{compared to the standard implementation of the Navier slip condition}. The idea is to apply \changeB{a given} slip length with respect to a \emph{virtual} ``staggered'' boundary located in between the physical boundary and the location of the face-centered velocity nodes (see Fig.~\ref{fig:staggered-slip}). \changeC{Mathematically, the distance $\tilde{L}$ to the virtual boundary can be expressed as}
\begin{align} 
\tilde{L} =  L + \frac{\omega \Delta x}{2}, 
\end{align}
\changeC{where $L$ is the distance to the physical boundary.} Consequently, the applied ghost velocity is 
\begin{align}
\label{eqn:ghost_velocity_staggered_slip}
\changeC{v_{\text{ghost}} = v \ \frac{2 \tilde{L} - \Delta x (1+\omega)}{2 \tilde{L} + \Delta x (1-\omega)}, \quad \omega \in [0,1]}.
\end{align}
\changeB{Note that here we keep $\tilde{L}$ fixed at the (physically) prescribed value when refining the mesh. The virtual boundary allows to apply larger ``counter velocities'' in the ghost cell layer than the standard Navier slip or no-slip conditions. Through (discrete) viscous forces this leads to a stronger damping or numerical dissipation. The quantity $\omega \in [0,1]$ is an empirical parameter that allows to control the amount of damping (see Appendix~\ref{section:staggered_slip_appendix}). For a staggered slip length $\tilde{L}$ smaller than $\omega \Delta x /2$, the staggered slip can be interpreted as standard Navier slip with a \emph{negative} slip length, while $\tilde{L} = \omega \Delta x/2$ corresponds to $L=0$ (numerical slip).} Note that a small staggered slip length may lead to very small numerical timesteps since the velocity $v_{\text{ghost}}$ in the ghost cells may become very large. Therefore, one cannot choose arbitrary small $\tilde{L}$ in practice.\\
\\
In the numerical simulations of the breakup process, we observed \changeB{spurious velocities at the contact line on the hydrophilic stripe} leading to capillary waves propagating along the droplet. The staggered slip condition allows to significantly dampen these unphysical velocities at the contact line. Moreover, it is found that the breakup dynamics of the capillary bridge is unaffected by the choice of $\omega$; see Appendix~\ref{section:staggered_slip_appendix} for more details. Hence, the staggered slip condition \changeC{only serves} as a stabilization method in the present study. \changeB{Unless stated otherwise, we choose $\omega=2/3$ and $\tilde{L} = 500 \, \text{nm}$ in the numerical simulations. See Appendix~\ref{section:staggered_slip_appendix} and \cite{Fricke2021} for more information on the staggered slip condition and its effect on dynamic wetting simulations.}

\makeatletter
\begin{figure}[ht]
 \subfigure[]{
 {
 \footnotesize
    \def\svgwidth{.55\textwidth}
      \begingroup \makeatletter \providecommand\color[2][]{\errmessage{(Inkscape) Color is used for the text in Inkscape, but the package 'color.sty' is not loaded}\renewcommand\color[2][]{}}\providecommand\transparent[1]{\errmessage{(Inkscape) Transparency is used (non-zero) for the text in Inkscape, but the package 'transparent.sty' is not loaded}\renewcommand\transparent[1]{}}\providecommand\rotatebox[2]{#2}\newcommand*\fsize{\dimexpr\f@size pt\relax}\newcommand*\lineheight[1]{\fontsize{\fsize}{#1\fsize}\selectfont}\ifx\svgwidth\undefined \setlength{\unitlength}{922.55406658bp}\ifx\svgscale\undefined \relax \else \setlength{\unitlength}{\unitlength * \real{\svgscale}}\fi \else \setlength{\unitlength}{\svgwidth}\fi \global\let\svgwidth\undefined \global\let\svgscale\undefined \makeatother \begin{picture}(1,0.65418633)\lineheight{1}\setlength\tabcolsep{0pt}\put(0,0){\includegraphics[width=\unitlength,page=1]{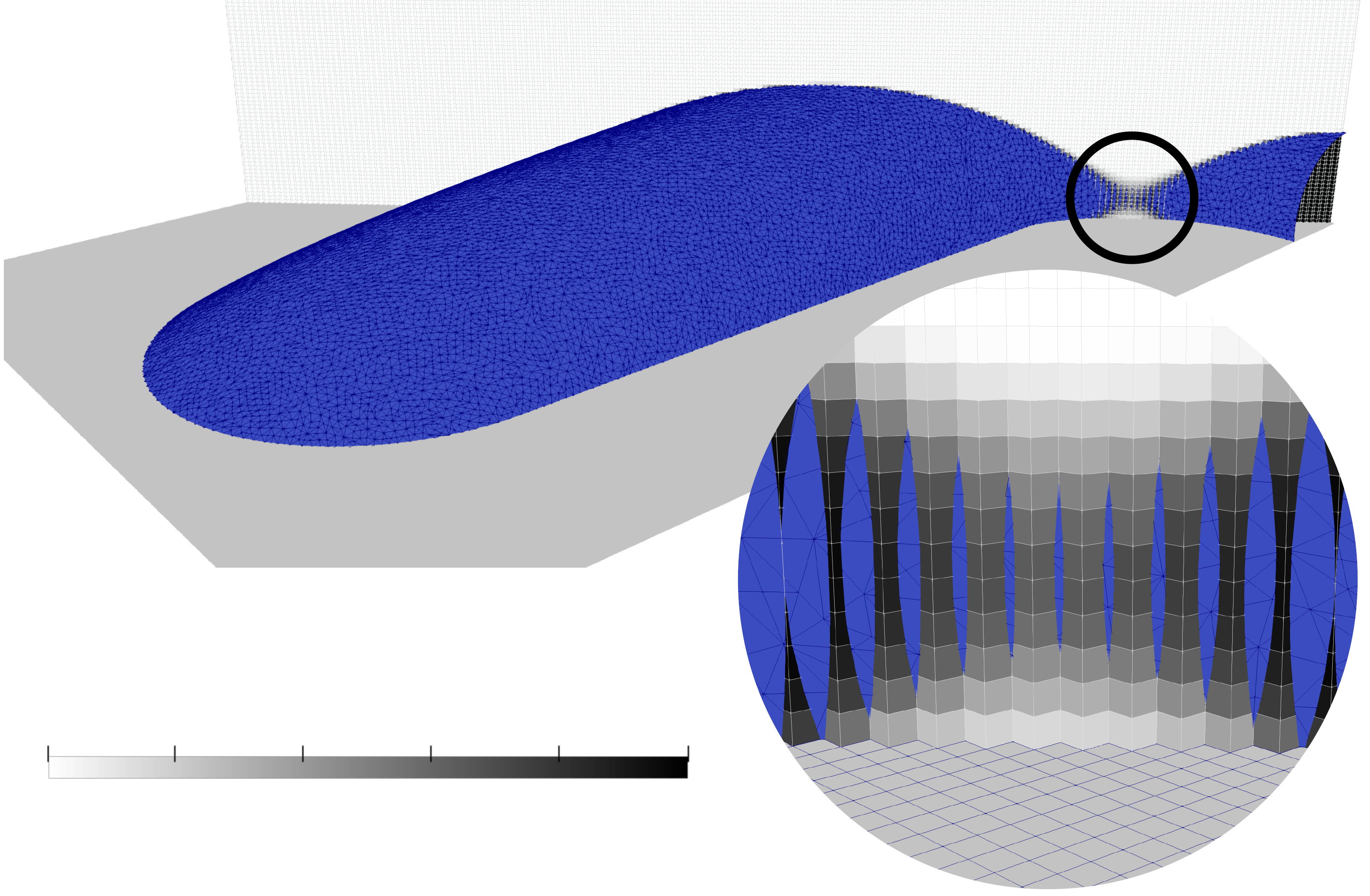}}\put(0.24809907,0.15061447){\makebox(0,0)[lt]{\smash{\begin{tabular}[t]{l}$f$\end{tabular}}}}\put(0.01718314,0.11919682){\makebox(0,0)[lt]{\smash{\begin{tabular}[t]{l}$0.0$\end{tabular}}}}\put(0.47668818,0.11919682){\makebox(0,0)[lt]{\smash{\begin{tabular}[t]{l}$1.0$\end{tabular}}}}\put(0.10865048,0.11919682){\makebox(0,0)[lt]{\smash{\begin{tabular}[t]{l}$0.2$\end{tabular}}}}\put(0.19750036,0.11919682){\makebox(0,0)[lt]{\smash{\begin{tabular}[t]{l}$0.4$\end{tabular}}}}\put(0.29371909,0.11919682){\makebox(0,0)[lt]{\smash{\begin{tabular}[t]{l}$0.6$\end{tabular}}}}\put(0.38299895,0.11919682){\makebox(0,0)[lt]{\smash{\begin{tabular}[t]{l}$0.8$\end{tabular}}}}\put(0,0){\includegraphics[width=\unitlength,page=2]{figures/drop_breakup_1.pdf}}\end{picture}\endgroup        \label{fig:conversion_overview}
 } 
 }
 \hspace{0.05\textwidth}
 \subfigure[]{
 {
 \footnotesize 
    \def\svgwidth{0.3\textwidth}
    \begingroup \makeatletter \providecommand\color[2][]{\errmessage{(Inkscape) Color is used for the text in Inkscape, but the package 'color.sty' is not loaded}\renewcommand\color[2][]{}}\providecommand\transparent[1]{\errmessage{(Inkscape) Transparency is used (non-zero) for the text in Inkscape, but the package 'transparent.sty' is not loaded}\renewcommand\transparent[1]{}}\providecommand\rotatebox[2]{#2}\newcommand*\fsize{\dimexpr\f@size pt\relax}\newcommand*\lineheight[1]{\fontsize{\fsize}{#1\fsize}\selectfont}\ifx\svgwidth\undefined \setlength{\unitlength}{2227.91363525bp}\ifx\svgscale\undefined \relax \else \setlength{\unitlength}{\unitlength * \real{\svgscale}}\fi \else \setlength{\unitlength}{\svgwidth}\fi \global\let\svgwidth\undefined \global\let\svgscale\undefined \makeatother \begin{picture}(1,1.00778582)\lineheight{1}\setlength\tabcolsep{0pt}\put(0,0){\includegraphics[width=\unitlength,page=1]{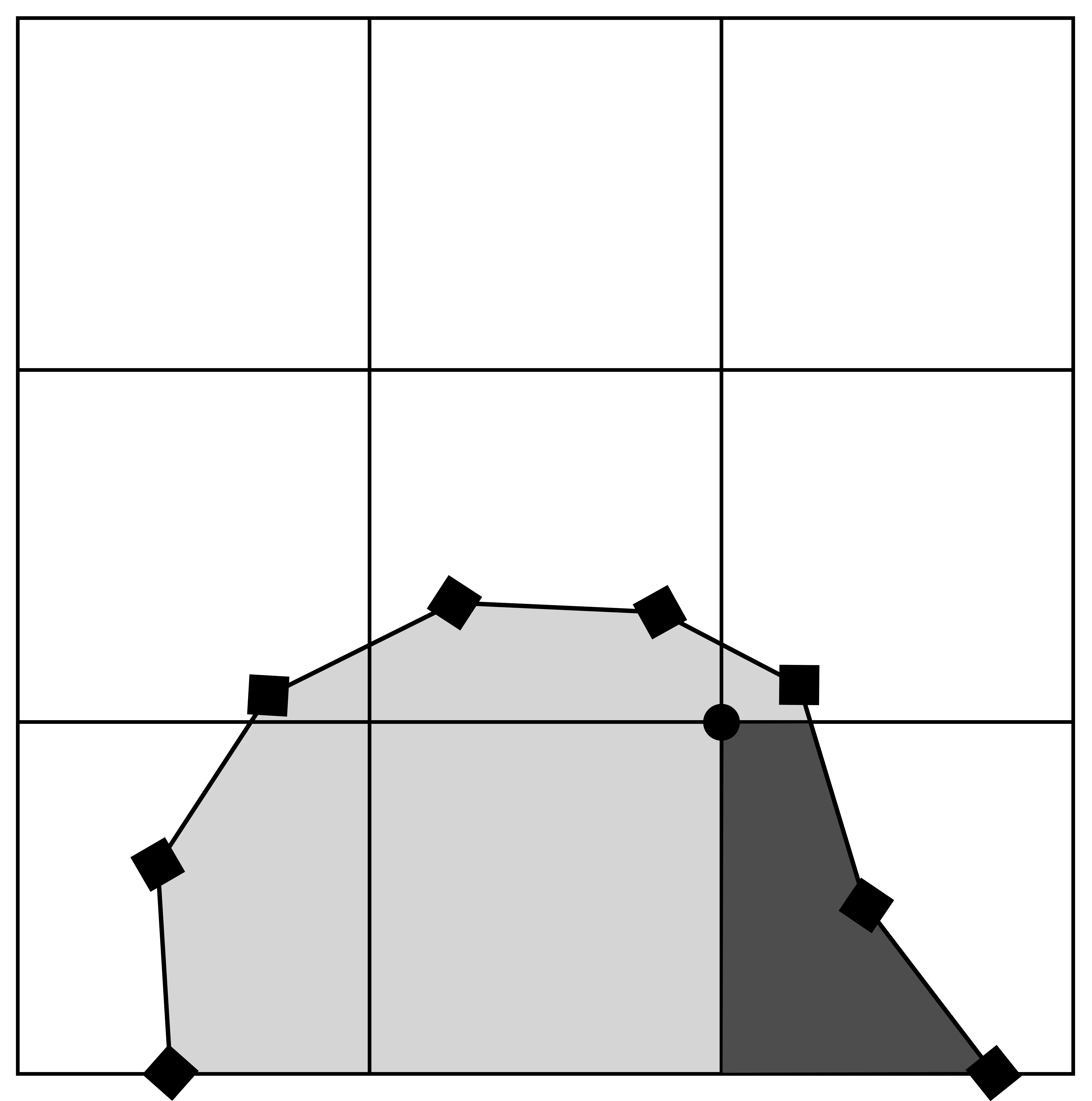}}\put(0.37429323,0.56684764){\makebox(0,0)[lt]{\smash{\begin{tabular}[t]{l}$\tilde{\Sigma}$\end{tabular}}}}\put(0,0){\includegraphics[width=\unitlength,page=2]{figures/intersected-cells-breakup-paper.pdf}}\end{picture}\endgroup      \label{fig:conversion_principle}
 }
 }  
 \caption{Initialization of the volume fraction field from STL-data. (a) Overview of the STL surface from Surface Evolver from Figure \ref{fig:surfaceEvolver_initGeo} (blue), together with the background mesh required for FS3D (greyscale). The enlarged image shows the STL surface with the resulting volume fraction in greyscale. (b) 2D illustration of the intersection between the background mesh (squares) and a volume bounded by the discrete surface $\tilde{\Sigma}$ (light grey) that may lead to non-convex intersection volumes (dark grey).}
\end{figure}
\makeatother

\paragraph{Computing volume fractions from the surface mesh given by Surface Evolver}

\changeB{As described in Section \ref{sec:minimal_surfaces}, the gas-liquid interface as shown in Figure \ref{fig:surfaceEvolver_initGeo} has been computed as a minimal surface using Surface Evolver. This tool provides the interface in the form of a triangulated surface (an STL-file) such as the blue triangulated surface shown in Figure \ref{fig:conversion_overview}. The dynamic breakup simulation using the VOF method requires the initial condition for the volume fraction defined in \eqref{eq:def_volume_fraction}. The computation of the initial volume fraction field $f$ uses the triangulated surface generated by Surface Evolver as indicated by the greyscale in the Figure \ref{fig:conversion_overview}.  Detailed information on the algorithm used for initializing volume fractions from surface meshes is available in \cite{Tolle2021vofi}, here we only briefly outline the algorithm.  The calculation of volume fractions from a surface mesh relies on the intersection of the surface mesh with each cell of the background mesh, as illustrated for the 2D case in Figure \ref{fig:conversion_principle}. Here, the background mesh points are indicated by circles, while squares indicate the mesh points of the STL surface. Arrows indicate the orientation of the STL surface.  Each intersected cell is decomposed into tetrahedra to simplify the geometrical intersection, shown as triangles in Figure \ref{fig:conversion_principle} in the cell with the phase-specific volume indicated with a darker gray color. A  sphere is defined for every tetrahedron in an intersected cell that encompasses the tetrahedron, schematically shown with the dashed line.  The tetrahedron centroid is its center, and the maximal Euclidean distance from the tetrahedron centroid to its corner points is the sphere radius. The tetrahedron sphere identifies a subset of triangles from the surface that intersects the tetrahedron. The phase-specific volume contained within the tetrahedron (black triangle in Figure \ref{fig:conversion_principle}), for which $\chi = 1$, is found by geometrically intersecting the tetrahedron with the triangulated surface subset. Therefore, the total phase-specific volume of a cell is the sum of the phase-specific volumes from all the tetrahedrons of the cell (total shaded volume in the cell in Figure \ref{fig:conversion_principle}). The volume fraction is then the ratio of the total phase-specific volume in the cell and the total cell volume. This algorithm returns a highly accurate geometrical value of ${0 < \chi < 1}$ for the intersected cells and its accuracy depends on the resolution of the surface mesh generated by the Surface Evolver.}
\begin{figure}[ht]
 \centering
 \includegraphics[width=0.4\textwidth]{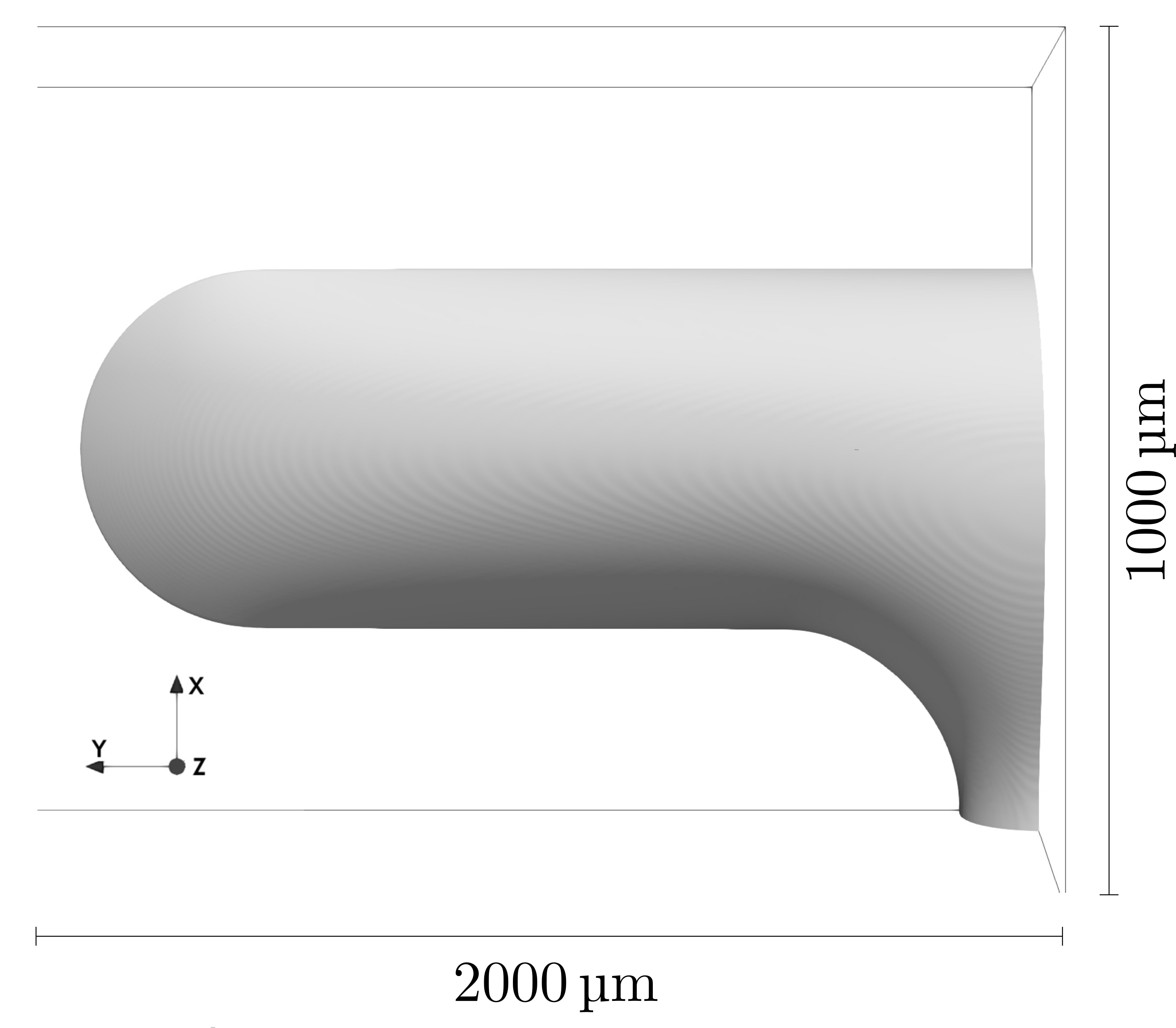}
 \caption{Setup for FS3D using symmetry conditions.}
 \label{fig:fs3d_setup}
\end{figure}

\paragraph{Numerical setup for FS3D:}
\changeC{The setup for FS3D is shown in Figure~\ref{fig:fs3d_setup}.} In order to save computational resources, we make use of the symmetry of the problem and simulate only a quarter of the droplet, while applying symmetry boundary conditions at the respective symmetry boundaries \changeC{($x=0$ and $y=0$)}. At the outer boundaries of the domain \changeC{(i.e., $x=x_{\text{max}}$, $y=y_{\text{max}}$ and $z=z_{\text{max}}$ )}, we apply no-slip for the velocity, which turns out to be irrelevant for the breakup dynamics (compared to, e.g., fixed pressure outflow boundary conditions). \changeC{The staggered Navier slip condition is applied at the solid boundary ($z=0$)}. The computational domain has a size of $1000 \mum \times 2000 \mum \times 500 \mum$ \changeC{in the $x$-, $y$- and $z$-direction} and is subdivided into an equidistant mesh of $2N \times 4N \times N$ cells, where $N$ is varied between $48$ and $128$. Therefore, the maximum resolution of the mesh is $500 \mum /128 \approx 3.9 \mum$ in each direction. Unfortunately, adaptive mesh-refinement is currently not available for our version of FS3D. The physical parameters used in the numerical simulation are listed in Table~\ref{tab:physicalParams}.

\begin{table}[ht]
	\centering
	\begin{tabular}{ c| c| c| c| c}
		$\rho_l$ [$\text{kg}/\text{m}^3$] & $\rho_g$ [$\text{kg}/\text{m}^3$] & $\eta_l$ [$10^{-6} \frac{\text{kg}}{\text{m}\cdot\text{s}}$] & $\eta_g$ [$10^{-6} \frac{\text{kg}}{\text{m}\cdot\text{s}}$] & $\sigma$ [$\frac{\text{mN}}{\text{m}}$] \\
		\hline
		$997.05$ & $1.17$ & $890.45$ & $18.5$ & $71.96$
	\end{tabular}
	\caption{Physical parameters for the numerical simulations (\changeC{literature values} for $T=298 \, \text{K}, \, p=100 \, \text{kPa}$).}
	\label{tab:physicalParams}
\end{table}
 
\section{Results and Discussion}
\label{section:results}

\subsection{Qualitative Comparison between Experiments and Numerics}
\label{sec:qualitativeComparison}
\begin{figure}[ht]
 \centering
 \includegraphics[]{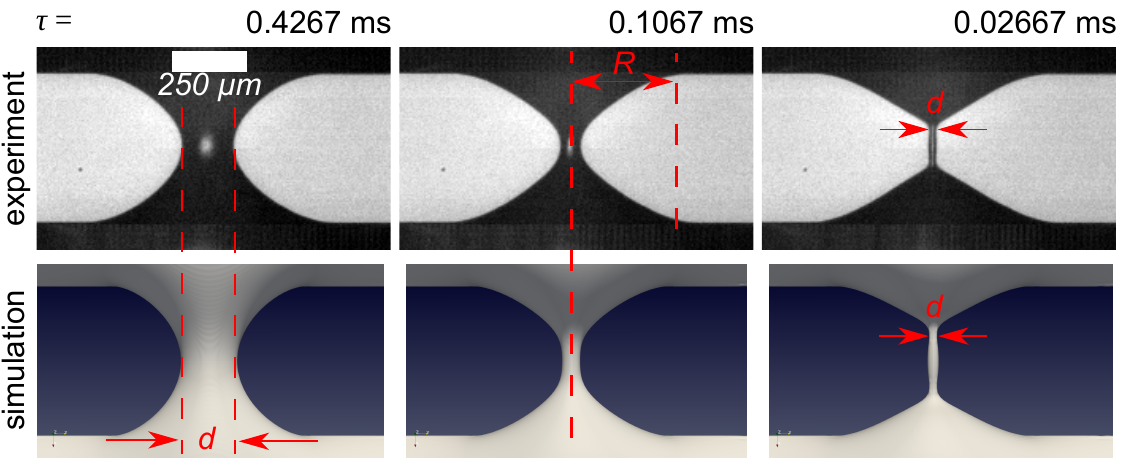}
 \caption{Qualitative comparison between experiment and simulation for $w_\mathrm{phil} = 500 \mum$ and $\alpha = 1$ at the same instants in time before breakup. The scale bar is valid for both experiment and simulation.} 
  \label{fig:qualitativeComparison}
\end{figure}

Since the focus of the present study is on the dynamics of the capillary bridge on the hydrophobic stripe, this part of the droplet is shown in more detail in Fig.~\ref{fig:qualitativeComparison}.
It shows the bridge at three different instants in time $\tau$ before the breakup event, i.e.\, $\tau = t_0 -t$, where $t_0$ is the breakup time, for both, experiment and simulation.
The geometrical parameters are $w_\mathrm{phil} = 500 \mum$ and $\alpha = 1$.
In the top row, the black and dark gray regions represent the liquid (with a reflection of the light in the middle of the capillary bridge), while the liquid is white in the bottom row.
The images in both rows also show some liquid that is wetting the hydrophilic stripes in the top and bottom part of each frame.

Qualitatively, in both experiments and simulations, the capillary bridge develops from a catenoid type to a narrower shape.
Then a liquid thread is formed that is getting constricted at two points.
At $\tau$~=~0.4267~ms, which is close to the initial condition for the simulations, the minimal width $d$ of the capillary bridge in both simulation and experiment is approximately the same.
This is also true for smaller $\tau$ until a liquid thread forms in the final instants before breakup at $\tau$~ =~0.02667~ms, when the minimum width of the bridge, as well as the bulge that forms in the middle, is larger in the simulations.
Note that $d$ is initially located in the middle of the capillary bridge.
After the elongated thread has formed, the position of the minimum width moves away from the center, as indicated at $\tau$~ =~0.02667~ms in Fig.~\ref{fig:qualitativeComparison}.
Below we will show that this is the starting point of a second dynamic regime.
Besides the bulge and the minimal width $d$, also the length of the thread differs between experiment and simulation.
The width of the capillary bridge at the instance when it becomes unstable is of the order of 200~$\mum$.
On this length scale, unavoidable contaminations can lead to significantly smaller contact angles than the measured receding contact angle (see e.g. Park et al.~\cite{Park2012} and the supplementary material of Hartmann and Hardt~\cite{Hartmann2019}).
These phenomena can neither be captured in the simulations nor in the calculation of the initial droplet configuration with Surface Evolver.
This might explain the deviations in the initial configuration between experiments and simulations, for example, the slightly bigger curvature of the capillary bridge in the plane parallel to the surface which is obvious at $\tau$~=~0.4267~ms.
Also the absolute value of the parameter $R$, which is the maximum width of the capillary bridge at the boundary between the hydrophobic and the hydrophilic stripe, differs slightly between experiment and simulation, though it stays constant in both.
\\
Because not all of the influencing factors relevant in the experiments could be incorporated in the simulations, quantitative agreement between experiments and simulations is not expected in all aspects of the static and dynamic wetting phenomena observed.
For quantitative comparison, the focus, therefore, lies on one specific aspect, namely the breakup dynamics of the liquid bridge, i.e. the time evolution of the narrowest section of the bridge. 
\subsection{Phase Space Representation of the Breakup Dynamics}
Following the literature, the breakup dynamics is usually described via the minimum width $d$ as a function of the time $\tau$ before the breakup event, i.e.\, $\tau = t_0 -t$, where $t_0$ is the breakup time. For the inviscid breakup of a free capillary bridge, it can be shown by means of an asymptotic analysis that this function follows the power law \eqref{eqn:inviscid_breakup_power_law}. However, the precise time of the breakup event is hard to determine both in experiments (due to finite spatial and temporal resolution) and in the simulation. Note that the choice of $t_0$ can have a large effect on the effective exponent that is extracted from the data. It has been reported that the same set of data appears to be represented by power laws $d(\tau) \propto \tau^\nu$ with an exponent $\nu$ ranging from $0.6$ to $0.8$ depending on the choice of $t_0$ \cite{Burton2007}. Both Li and Sprittles \cite{Li2016} and Deblais et al.\ \cite{Deblais2018} eliminated the dependency on the breakup time by rewriting the power law \eqref{eqn:inviscid_breakup_power_law} according to
\begin{align} 
\label{eqn:scaling_deblais_et_al}
d = C \left(\frac{\sigma}{\rho}\right)^{1/3} \, (t_0 -t)^{2/3} \quad \Leftrightarrow \quad d^{3/2} = C^{3/2} \left(\frac{\sigma}{\rho}\right)^{1/2} (t_0 -t). 
\end{align}
\changeC{Hence, relation \eqref{eqn:inviscid_breakup_power_law} holds if and only if} the quantity $d^{3/2}$ is linear in time and, \changeC{in this case}, the value of $C$ can be found from the slope \changeC{(i.e.\ the time derivative)} of $d^{3/2}$. \\

\begin{figure}[ht]
\subfigure[$t_0 = 0.632 \, \text{ms}$, $\nu = 2/3$.]{\includegraphics[width=8cm]{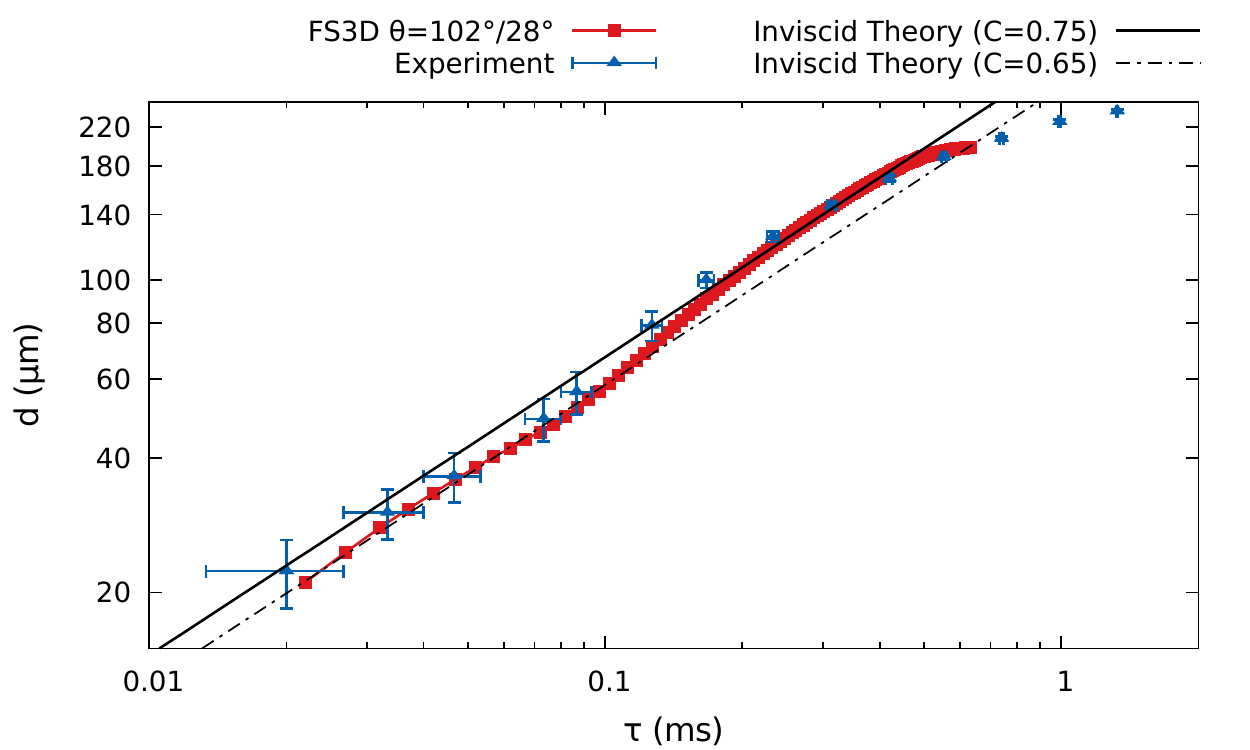}\label{fig:sensitivity_v_2_3}}
\subfigure[$t_0 = 0.622 \, \text{ms}$, $\nu = 0.4755$.]{\includegraphics[width=8cm]{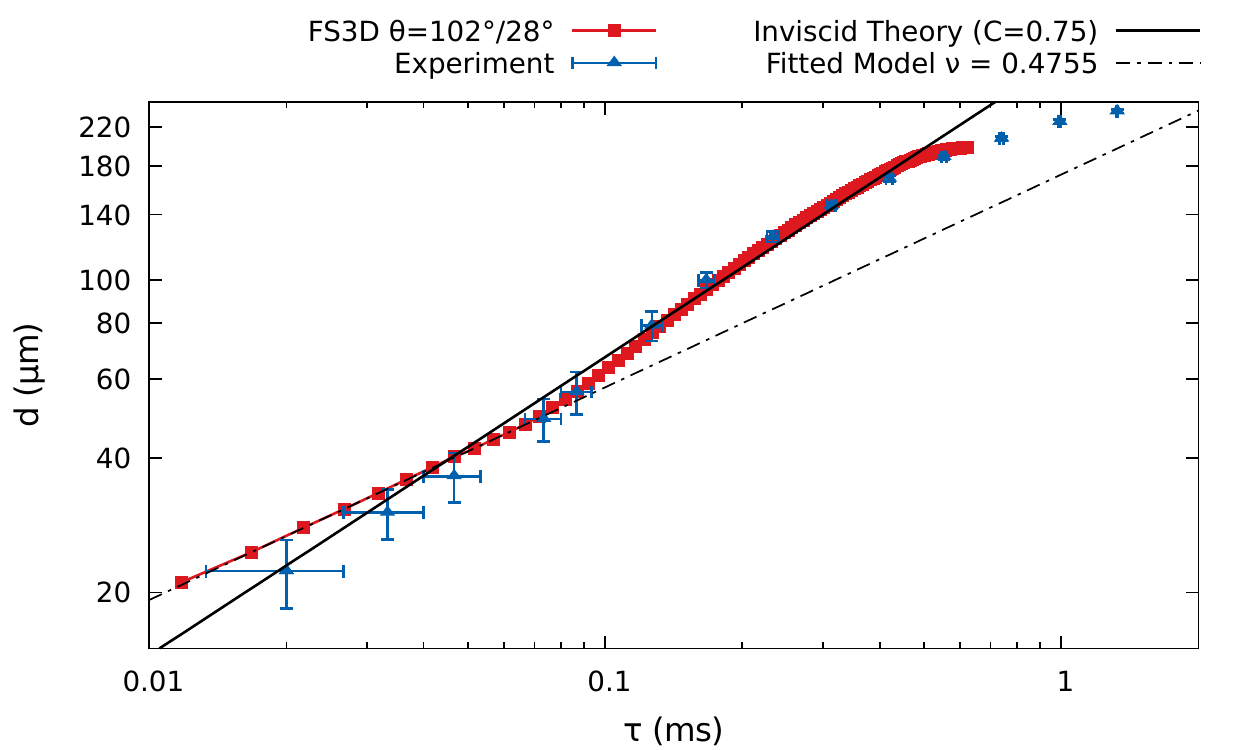}\label{fig:sensitivity_v_0_5}}
\caption{Breakup dynamics for different choices of the breakup time $t_0$ for the numerical data.}
\label{fig:sensitivity_breakup_time}
\end{figure}

In the Volume-of-Fluid simulation, the actual breakup is usually mesh-dependent since it is ultimately performed by the interface reconstruction algorithm. Moreover, since the breakup process involves very small length scales, it cannot be fully resolved by the numerics. Therefore, the numerical results can only be considered meaningful down to a certain length scale determined by the computational mesh. In the present study, this length scale is approximately $20\, \mu m$. Consequently, the breakup time cannot be extracted from the numerics in a meaningful way without extrapolating the data.\\
\\
The experimental value of the breakup time is determined from the pictures taken by the high-speed camera.
The first image where the capillary bridge is pinched off defines the time $t^\star$ which is always larger than the real breakup time $t_0$.
Clearly, the breakup must occur between $t^\star$ and the time associated with the previous image.
We estimate the breakup time to be $t_0 = t^\star - \Delta t / 2$, where $\Delta t = 1.33 \cdot 10^{-2}$~ms follows from the frame rate of the high-speed camera which is 75,000 fps.
To account for the uncertainty in breakup time in the case of experiments, horizontal error bars with a total length of $\Delta t$ are drawn.
In the vertical direction, the error bar represents the standard deviation obtained from 5 experiments.\\
\\
A concrete example for the sensitivity of the simulation results with respect to the choice of the breakup time is given in Fig.~\ref{fig:sensitivity_breakup_time}. Besides the experimental results,  Fig.~\ref{fig:sensitivity_v_2_3} shows the simulation data for the choice $t_0 = 0.632 \, \text{ms}$. The latter value is found by fitting the data in the final regime $d \lesssim 50 \mum$ with the exponent $\nu = 2/3$. Fig.~\ref{fig:sensitivity_v_0_5} shows the same numerical data set for a slightly smaller breakup time \changeD{$t_0 = 0.622 \, \text{ms}$}. In this case, the numerical data agrees well with a power law with a smaller exponent \changeD{$\nu \approx 0.48$}, which has been obtained from the systematic phase space analysis described below. Note that the difference between the two choices \changeD{$\Delta t_0 = 10 \, \mu\text{s}$} is smaller than the inverse frame rate of the high-speed camera which is approximately $\Delta t = 13 \, \mu\text{s}$. This example clearly shows the need for a systematic method that does not rely on the breakup time.

\paragraph{Phase space representation.} We apply a different approach to describe the breakup dynamics which is \emph{independent} of the choice of the breakup time and allows to identify different dynamic regimes in a systematic way. \changeC{The basic idea is to use the variables $(d, -\dot{d})$ rather than $(\tau, d)$ to describe the evolution of the minimum width.} This approach was first applied by Li and Sprittles~\cite{Li2016} and is systematically elaborated in the following.\\
\\
Given the minimum width of the capillary bridge as a function of physical time $t$, we consider the \emph{breakup speed}, i.e.\ the time derivative
\[ V = -\dot{d} \]
as a function of the \emph{minimum width} itself, i.e.\ we formally define
\begin{align}\label{eqn:space_space_definition}
V(\tilde{d}):=-\dot{d}(\tilde{t}), \quad \text{where} \quad \tilde{t} := d^{-1}(\tilde{d}).
\end{align}
The latter quantity is well-defined since the minimum width $d$ is a monotonically decreasing function (over the time scale of the breakup process). Obviously, the function $V(d)$ is invariant with respect to shifts in the time coordinate. \changeD{Moreover, it can be shown mathematically that the remaining information of the function $d(\tau)$ is still contained in $V(d)$. Indeed, the function $d(\tau)$ can be reconstructed from $V(d)$ up to a shift in time; see \cite{Fricke2021} for details.}\\
\\
For the power law \eqref{eqn:inviscid_breakup_power_law} describing the inviscid regime we have
\[ d(t) = C \left(\frac{\sigma (t_0-t)^2}{\rho}\right)^{1/3} \quad \Rightarrow \quad \dot{d}(t) = - \frac{2}{3} C \left(\frac{\sigma}{\rho (t_0-t)}\right)^{1/3} = -\frac{2}{3} C^{3/2} \left(\frac{\sigma}{\rho \, d(t)}\right)^{1/2}.   \]
Hence the power law \eqref{eqn:inviscid_breakup_power_law} translates to
\begin{align}
\label{eqn:inviscid_breakup_power_law_velocity}
V(d) = \frac{2}{3} C^{3/2} \left(\frac{\sigma}{\rho d}\right)^{1/2}.
\end{align}
More generally, one can easily show the relation 
\begin{align} 
d(t) = c (t_0-t)^\nu \quad \Rightarrow \quad V(d) = \nu c^{1/\nu} \, d^{1-1/\nu} = \tilde{c} \, d^{\tilde{\nu}},
\end{align}
which is valid for an arbitrary power law ($\nu, \, c > 0$). So the exponent $\tilde{v}$ and the prefactor $\tilde{c}$ obtained from the phase space diagram can be transformed into the standard representation via
\begin{align}
\nu = \frac{1}{1-\tilde\nu}, \quad c = \left(\frac{\tilde{c}}{\nu}\right)^\nu = [(1-\tilde\nu) \, \tilde{c}]^{1/(1-\tilde\nu)}.
\end{align}
For the material parameters of water in air (temperature $T = 298 \, \text{K}$, pressure $p = 100 \, \text{kPa}$, see Table~\ref{tab:physicalParams}), the relations \eqref{eqn:inviscid_breakup_power_law} and \eqref{eqn:inviscid_breakup_power_law_velocity} for a free liquid bridge take the form
\begin{align*}
d(\tau) = C \cdot 416 \, \hat{\tau}^{2/3} \, \mum, \quad V(d) = C^{3/2} \cdot 5657 \, \hat{d}^{-1/2} \, \frac{\mum}{\text{ms}},
\end{align*}
where $\tau = \hat{\tau} \, \text{ms}$ and $d = \hat{d} \mum$. In the present paper, these relations will be referred to as ``inviscid theory''.\\
\\
Note that the above method requires to differentiate potentially noisy data with respect to time.
This issue has been addressed via filtering out high-frequency oscillations in the experimental values of $V(d)$ by locally fitting a straight line to the data (using six neighboring points).
Despite this difficulty, the method allows studying the breakup dynamics in detail without the uncertainty in choosing $t_0$. In the following, we will report both $V(d)$ and $d(t_0-t)$ for completeness. \subsection{Quantitative Comparison between Experiments and Simulations}
\label{sec:quantComparison}
\subsubsection{The case $\alpha=1$:}
\begin{figure}[ht]
\centering
\includegraphics[width=10cm]{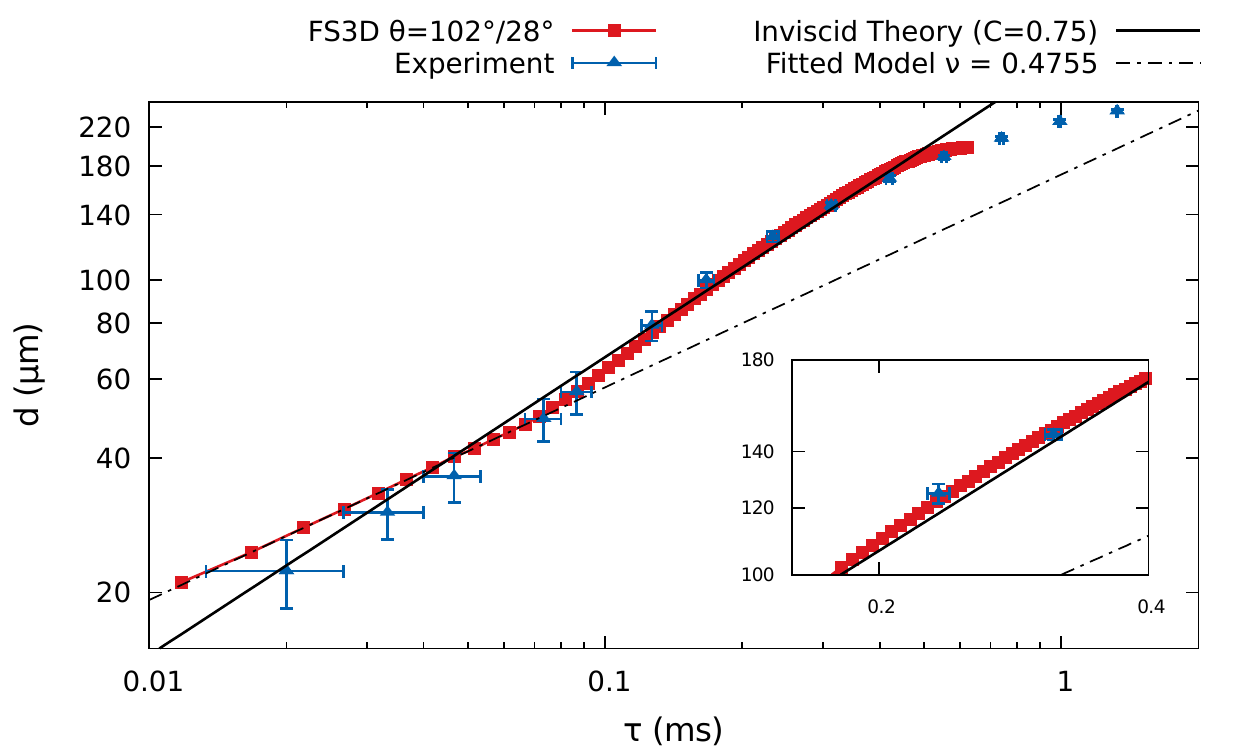}
\caption{Minimum bridge width as a function of time before breakup, experiment vs.\ simulation ($\alpha = 1$).}
\label{fig:bridge_diameter_dynamics_exp_vs_simulation_alpha1}
\end{figure}

We first consider the case $\alpha = 1$. Figure~\ref{fig:bridge_diameter_dynamics_exp_vs_simulation_alpha1} compares the experimental data with numerical simulations for $\thetaphob=102^\circ$ and $\thetaphil=28^\circ$. Before the onset of breakup process ($d \gtrsim 180 \mum$), the bridge width in the experiments is decreasing due to evaporation. This continues until the critical width is reached, indicated by the starting point of the simulation curve in Fig.~\ref{fig:bridge_diameter_dynamics_exp_vs_simulation_alpha1} corresponding to the configuration computed by Surface Evolver.\\
In the region between about $100 \mum$ and $180 \mum$, the numerical results are in good quantitative agreement with the experimental data and the inviscid theory for $C \approx 0.75$. However, a careful inspection shows that the data in this region do not exactly lie on a straight line in the double logarithmic diagram. Indeed there is a non-vanishing curvature visible.
Below approximately $100 \mum$, there is a transition to a second dynamic regime that follows a power law with an exponent $\nu$ close to 0.5. This value is significantly smaller than the one for a free capillary bridge in the inviscid regime. Note that the exponent is obtained from fitting the numerical data in the phase space diagram, see below.\\

\begin{figure}[ht]
\centering
\includegraphics[width=10cm]{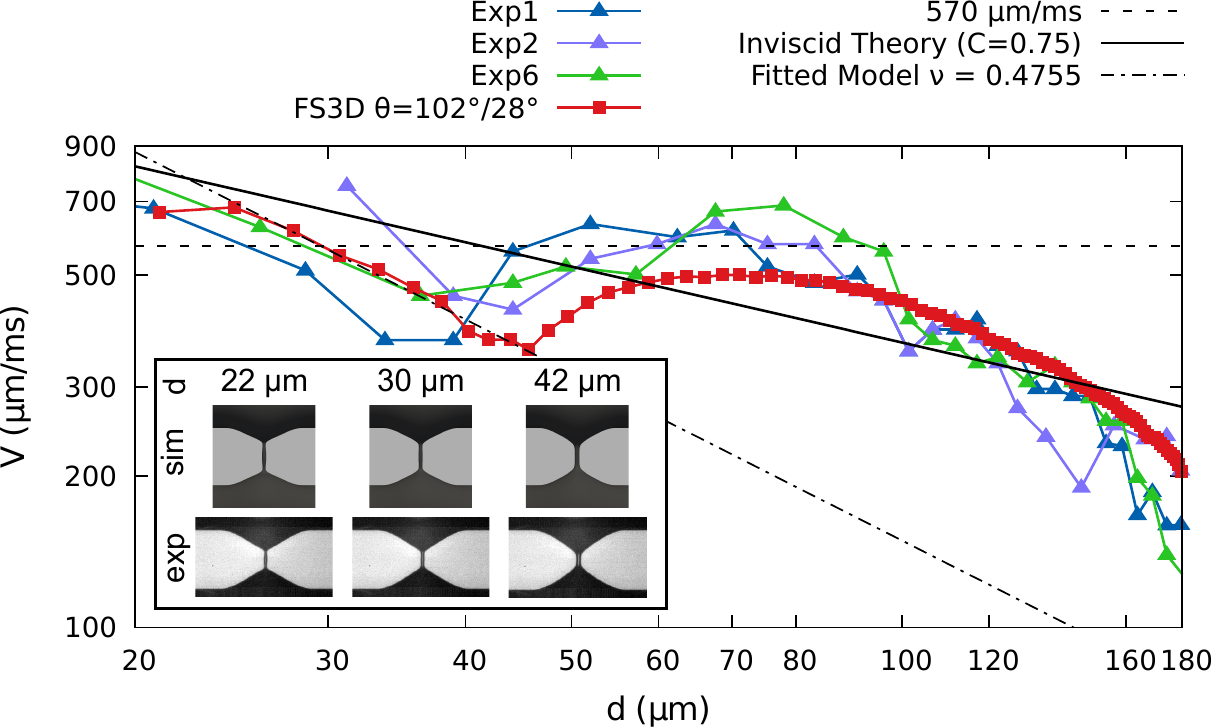}
\caption{Breakup speed ($\alpha = 1$) as a function of the bridge width compared to 3 different experiments. The pictures inside of the graph show the (closest) images for simulations (sim) and experiments (exp) to the minimum bridge width indicated in the top row.}
\label{fig:alpha_1_breakup_velocity}
\end{figure}

\paragraph{Phase space representation:} Figure~\ref{fig:alpha_1_breakup_velocity} shows the same data set in the phase space diagram. This representation of the data allows a much more detailed study of the breakup process independent of the breakup time. Note that the experimental data refer to three \emph{individual} repetitions of the same experiment. By considering individual experiments rather than averaged quantities, we are able to visualize detailed features of the dynamics that might otherwise be averaged out. Moreover, note that the noise in the experimental data for the bridge width translates to a noise in the calculated breakup speed. This explains the oscillations in the experimental data which are not present in the numerical data.\\
\\
The phase space diagram in Fig.~\ref{fig:alpha_1_breakup_velocity} shows that the dynamics in the region $100 \mum$ and $180 \mum$ agrees with the inviscid theory for $C \approx 0.75$ only in an average sense. In fact, there is no clear scaling relation in the latter region.
At approximately $70\mum$, the speed in the experiment reaches a local maximum of approximately $570 \mum/\text{ms}$ on average (indicated by the dashed horizontal line in Fig.~\ref{fig:alpha_1_breakup_velocity}) before it starts to decrease further towards a local minimum. This behavior is also clearly visible in the numerical data. We note that the position of the local minimum differs for different repetitions of the experiment (see Fig.~\ref{fig:alpha_1_breakup_velocity}). Note that the simulation predicts a smaller local maximum of approximately $500 \mum/\text{ms}$.\\ 

\begin{figure}
\centering
\includegraphics[width=0.5\columnwidth]{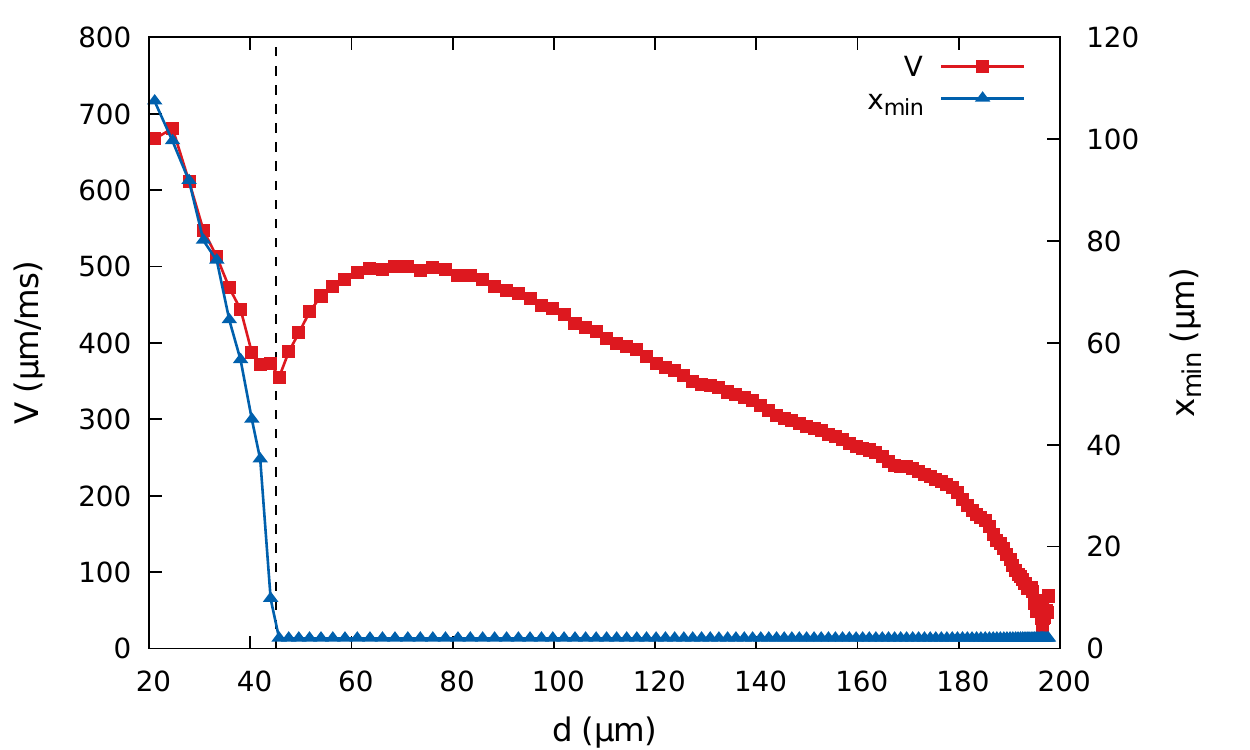}
\caption{Breakup speed $V$ and position of the local \changeC{minima $x_\text{min}$} (measured relative to the center of the hydrophobic stripe) in the numerical simulation ($\alpha=1$).}
\label{fig:motion_of_local_minimum}
\end{figure}

Moreover, the analysis of the numerical data shows that the local minimum in the breakup speed corresponds to the time instant in the process when the liquid filament has formed and the position of the minimum width \changeC{$x_{\text{min}}$} starts to move \changeC{symmetrically\footnote{The symmetry is enforced explicitly in the simulation but also observed in the experiment; see Fig.~\ref{fig:qualitativeComparison}.}} from the middle of the liquid bridge \changeC{($x=0$)} towards the \changeC{points} of the final pinch-off; see Fig.~\ref{fig:motion_of_local_minimum}. So, in fact, this can be understood as the starting point of a
second dynamic regime.
In the experiments, this final breakup behaves slightly differently. 
Though a filament has \changeC{formed at the time when the breakup speed reaches a local minimum}, the width of this filament first further decreases, while the filament has \changeC{approximately} the same width along its axis without any constriction point.
Compared to the simulations, the \changeC{location where the bridge has its minimum} starts to move \changeC{in a symmetrical manner} to the constriction points at a later instance in time, i.e. for a smaller value of $\tau$.\\
As can be seen from both the experimental and the numerical data in Fig.~\ref{fig:alpha_1_breakup_velocity}, the second dynamic process is not consistent with the inviscid theory for a free capillary bridge~\eqref{eqn:inviscid_breakup_power_law}.
In fact, by fitting the numerical data, the breakup velocity in the second breakup regime follows the power law
\begin{equation}
V(d) = \changeD{23879}\,\frac{\mum}{\text{ms}} \cdot \left(\frac{d}{\mum}\right)^{\changeD{-1.1029}}
\label{eqn:secDynFit_velo}
\end{equation}
which corresponds to 
\begin{equation}
d(\tau) = \changeD{171.95}\,\mum \cdot \left(\frac{\tau}{\text{ms}}\right)^{\changeD{0.4755}};
\label{eqn:secDynFit_dim}
\end{equation}
see Fig.~{\ref{fig:bridge_diameter_dynamics_exp_vs_simulation_alpha1}}. \changeD{Note that the last data point ($d \approx 21 \mum$) from the numerical simulation in Fig.~{\ref{fig:alpha_1_breakup_velocity}} has been excluded from the fit since the breakup speed appears to reach another local maximum in this region. Hence, the numerical data suggest that another regime transition may happen at smaller scales that cannot be resolved with the current approach.}
We attribute \changeD{the observed discrepancy to the inviscid scaling law} to the complex geometric configurations found in the second dynamic regime.
A simple scaling law such as~\eqref{eqn:inviscid_breakup_power_law} is only expected to work if the configuration of a system is essentially rescaled during its time evolution.
We do not expect that the scaling law to capture more complex scenarios such as a shift of the position of the minimum width of the filament. \changeC{In fact, the self-similarity of the interface shape, which is an essential assumption for the derivation of \eqref{eqn:inviscid_breakup_power_law} (see \cite{Keller1983}), is lost due to the presence of the solid substrate. Moreover,} as it has been pointed out by Deblais et al., ``the approach to asymptotic power laws can be slow and may pass through one or several transient regimes before the final universal regime is reached'' \cite{Deblais2018}. Hence the exponent $\nu$ may change on smaller length scales below the resolution of the present experiments and simulations.

\begin{figure}[ht]
 \subfigure[Breakup speed for different densities.]{\includegraphics[width=8cm]{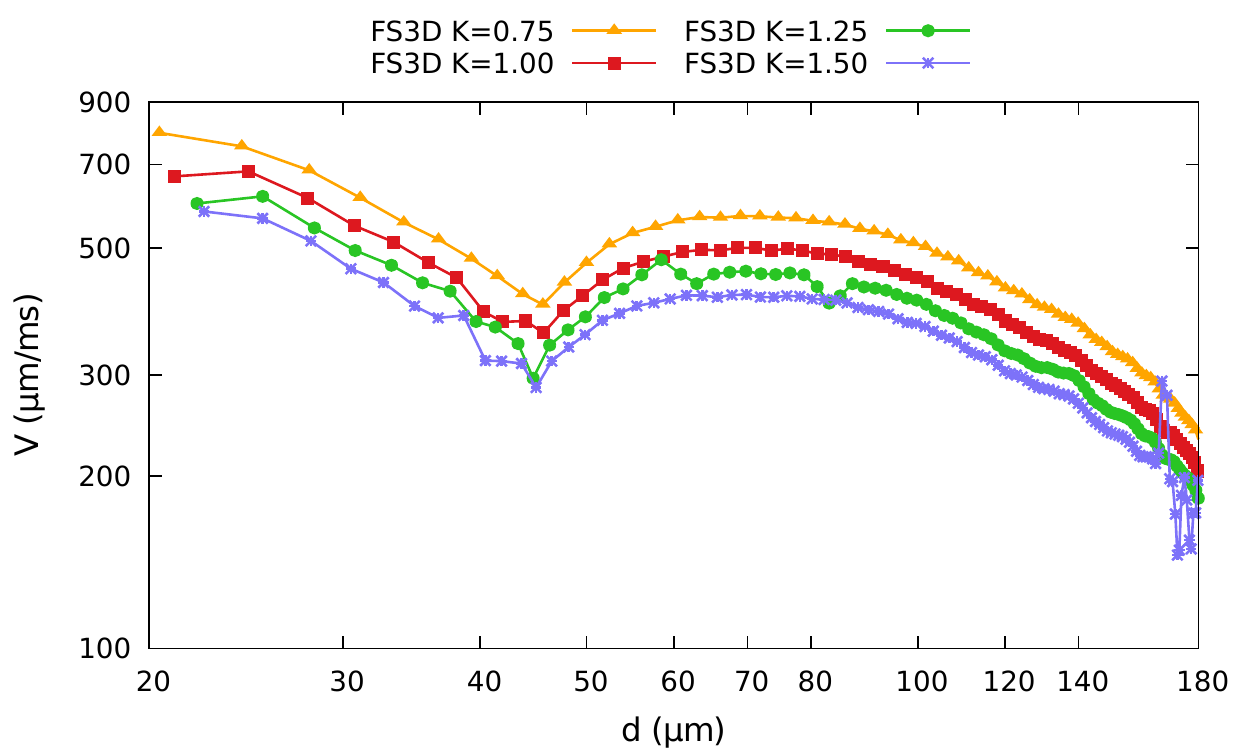}\label{fig:alpha_1_breakup_velocity_rho_study}}
 \subfigure[Dimensionless form of the breakup speed.]{\includegraphics[width=8cm]{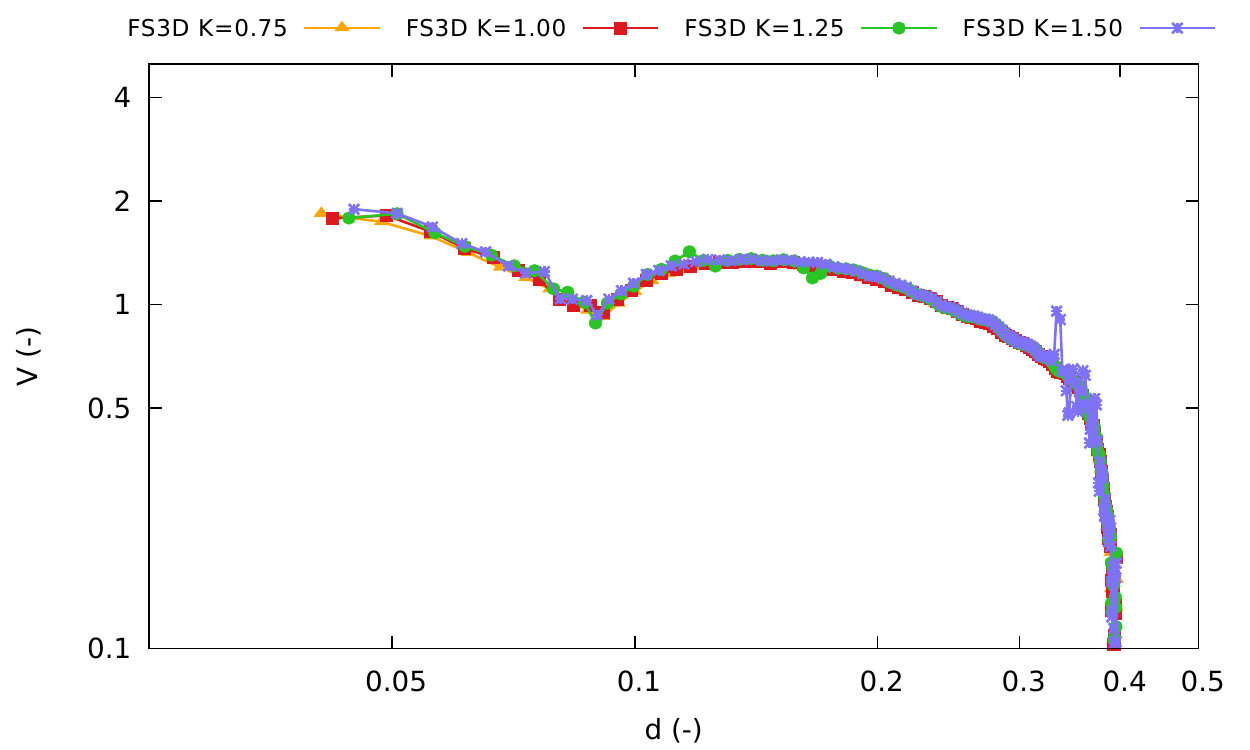}\label{fig:alpha_1_breakup_velocity_rho_study_non_dim}}
 \caption{\changeC{Variation of the liquid density and gas density for $\alpha=1$, $\thetaphob = 102^\circ$ and $\thetaphil = 28^\circ$.}}
 \label{fig:alpha_1_breakup_velocity_rho_study_summary}
\end{figure}
\changeC{\textbf{Influence of inertial forces:} We vary both the liquid density $\rho_l$ and the gas density $\rho_g$ in the simulation to study how inertial forces modify the breakup dynamics. Fig.~\ref{fig:alpha_1_breakup_velocity_rho_study} shows the breakup speed in the phase space diagram for four different values of the liquid and gas densities, i.e.\
\[ \rho_l = K \cdot 997.05 \ \text{kg}/\text{m}^3 , \quad \rho_g = K \cdot 1.17 \ \text{kg}/\text{m}^3 \quad \text{where} \quad K \in \{0.75, 1, 1.25, 1.5 \}. \]
As expected, the breakup speed decreases with increasing density. Remarkably, the data can be collapsed onto a single master curve by an appropriate nondimensionalization. Fixing a length scale $L$ in \eqref{eqn:inviscid_breakup_power_law} yields
\begin{align*}
\frac{d(\tau)}{L} = C \left( \frac{\sigma \tau^2}{\rho L^3} \right)^{1/3}.
\end{align*}
Hence, with the choice
\[ T = \sqrt{\frac{\rho L^3}{\sigma}} \]
for the time scale, equation \eqref{eqn:inviscid_breakup_power_law} takes the form
\begin{align}
\hat{d}(\hat{\tau}) = C \, \hat{\tau}^{2/3},
\end{align}
where $\hat{d} = d/L$ and $\hat{\tau} = \tau/T$. The corresponding velocity scale is
\begin{align}
V_{\text{scale}} = \frac{L}{T} = \sqrt{\frac{\sigma}{\rho L}} \propto \frac{1}{\sqrt{\rho}}.
\end{align}
Indeed, plotting the non-dimensional breakup speed $\hat{V} = V/V_{\text{scale}}$ for $$L=w_{\text{phob}} = 500 \mu \text{m}$$ collapses the data onto a single curve; see Fig.~\ref{fig:alpha_1_breakup_velocity_rho_study_non_dim}. This indicates that the observed process is indeed dominated by a balance of inertial and capillary forces despite the fact that the dynamics does \emph{not} simply follow the power law \eqref{eqn:inviscid_breakup_power_law}.\\
} 

\begin{figure}[ht]
 \subfigure[Mesh dependence study ($\tilde{L}=500 \nm$).]{\includegraphics[width=8cm]{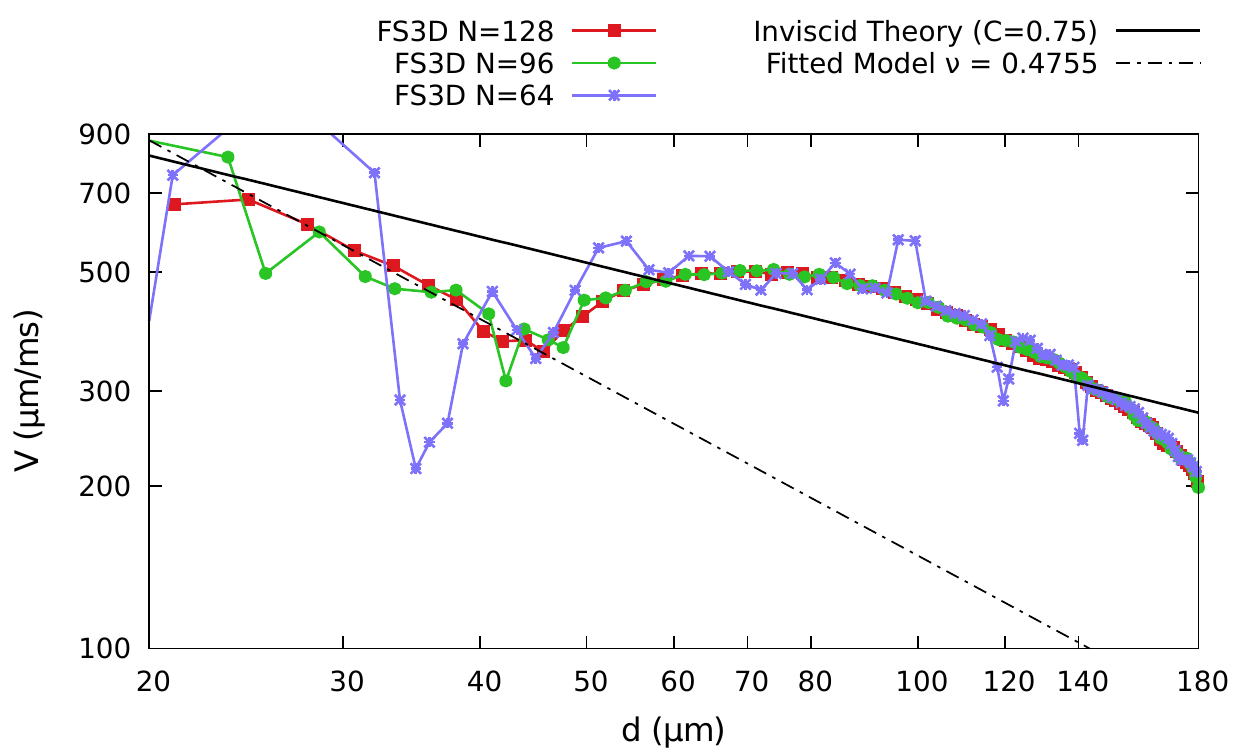}\label{fig:mesh_study_velocity_fs3d_alpha1}}
 \subfigure[Slip-length dependence study ($N=128$).]{\includegraphics[width=8cm]{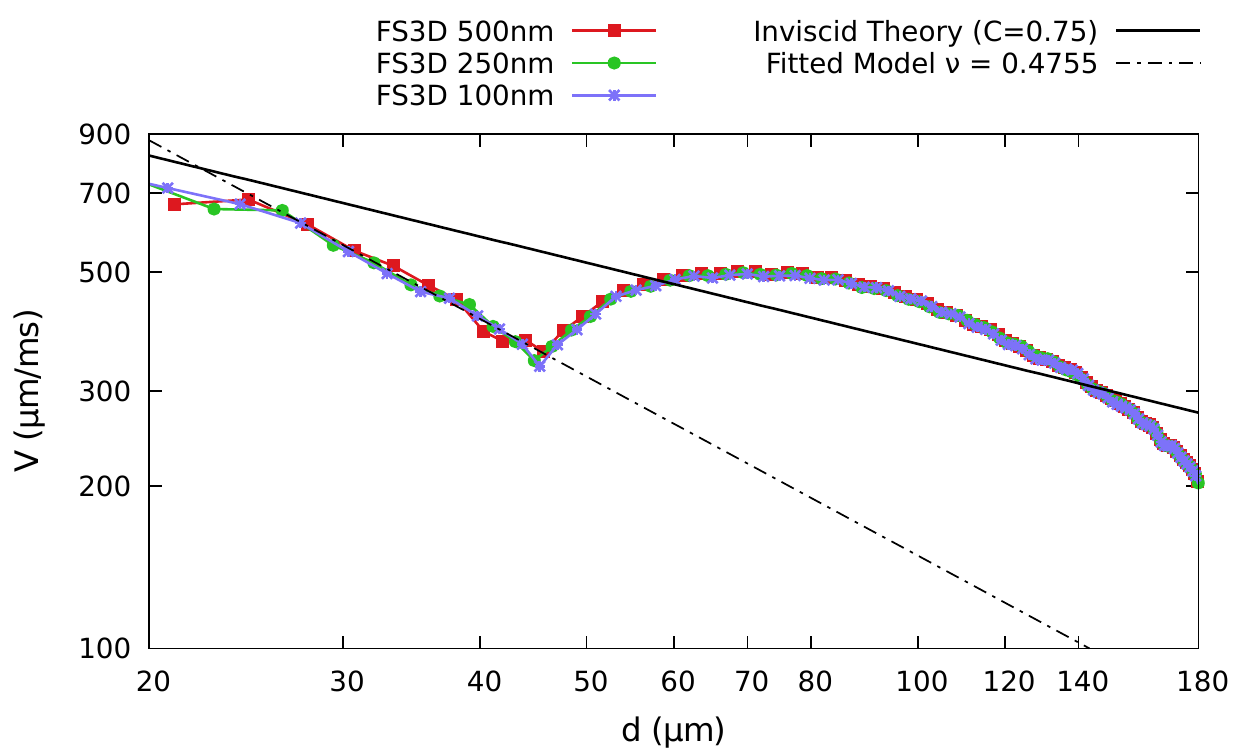}\label{fig:mesh_study_velocity_fs3d_staggered_slip_study_alpha1}}
 \caption{Mesh and slip-length dependence study for $\alpha=1$, $\thetaphob = 102^\circ$ and $\thetaphil = 28^\circ$.}
\end{figure}

\textbf{Mesh study:} To quantify the influence of the numerical discretization, Fig.~\ref{fig:mesh_study_velocity_fs3d_alpha1} shows numerical data at the breakup speed for different mesh resolutions ($8 N^3$ computational cells where $N \in \{64,96,128\}$). Besides some oscillations in the breakup speed on the coarse mesh, the results appear to be reasonably mesh-independent for $d$ larger than approximately $40 \mum$, whereas the simulation on the coarsest mesh ($N=48, \, \Delta x \approx 10.4 \mum$) appears to be under-resolved. The finest mesh ($N=128, \, \Delta x \approx 3.9 \mum$) delivers reasonable results for $d \gtrsim 10 \mum$. This corresponds to only $2.5$ computational cells within the bridge width. Since mesh convergence cannot be assured on this scale, the numerical data below approximately $20 \mum$ should not be used to draw a quantitative conclusion.\\

\begin{figure}[ht]
\subfigure[Viscosity variation.]{\includegraphics[width=8cm]{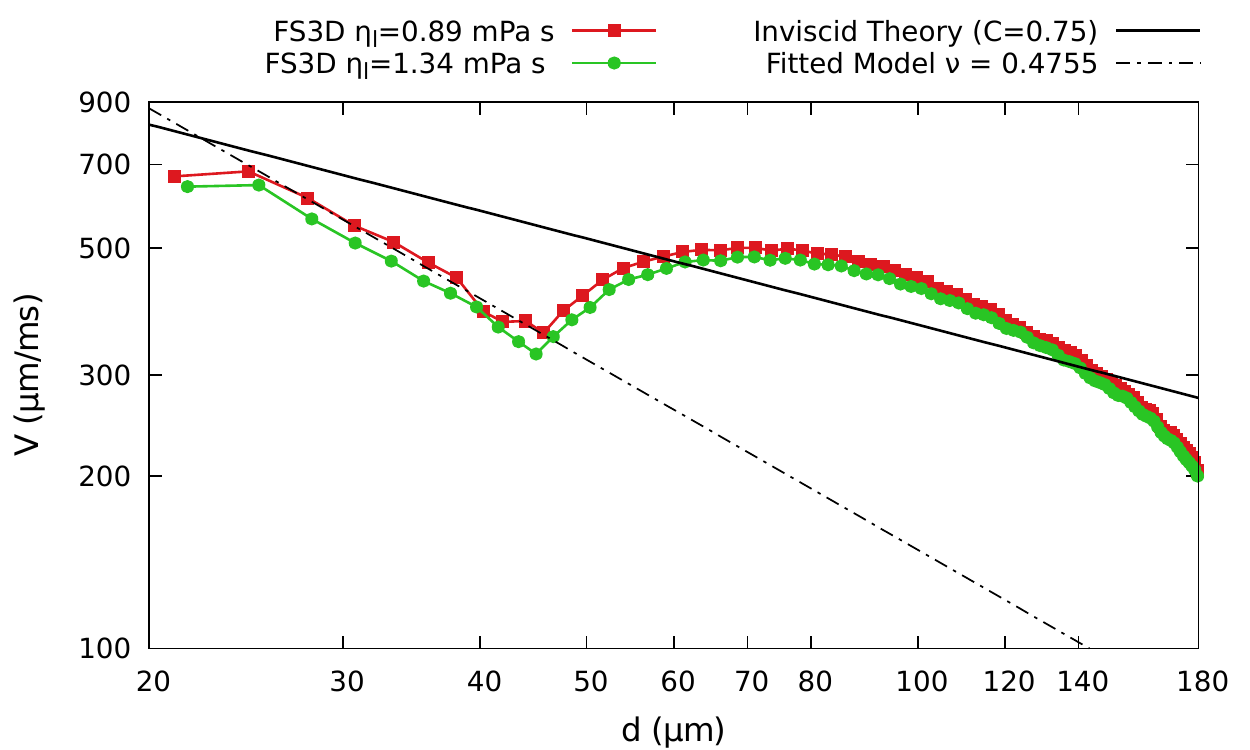}\label{fig:alpha_1_viscosity_study}}
\subfigure[Contact angle variation.]{\includegraphics[width=8cm]{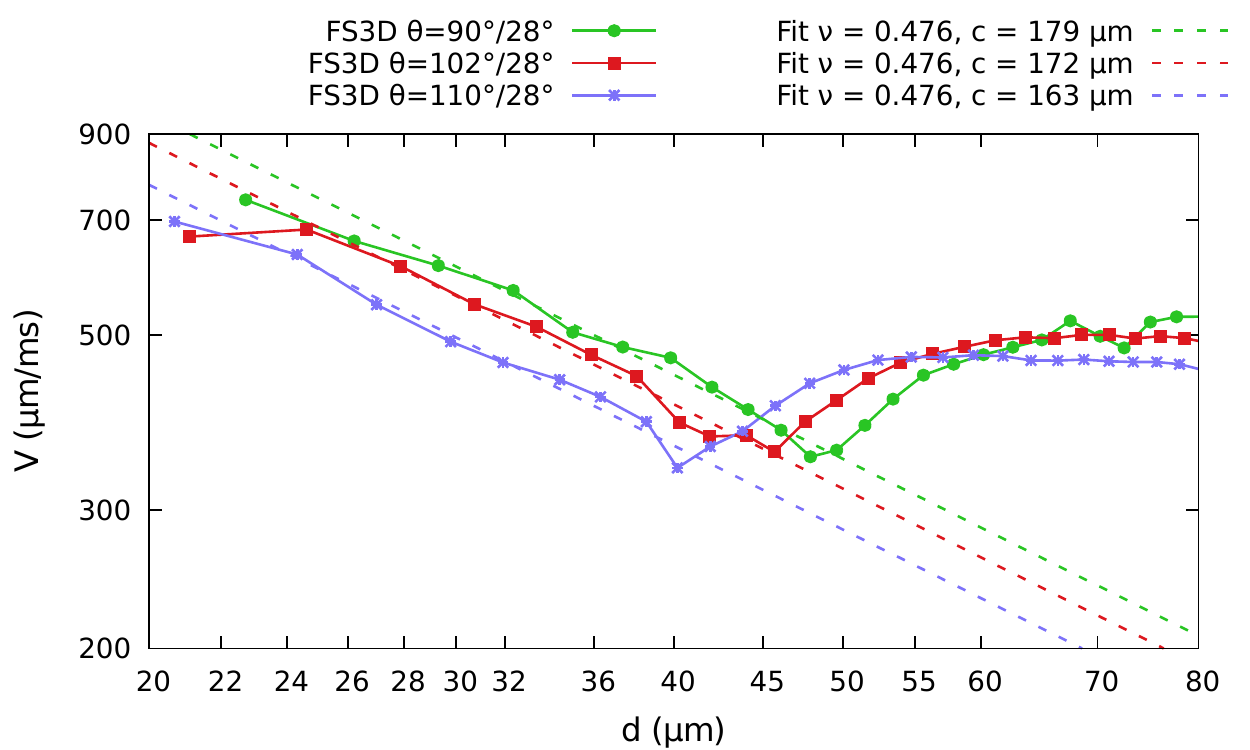}\label{fig:theta_study_alpha_1}}
\caption{Influence of the dynamic viscosity and the hydrophobic contact angle.}
\end{figure}

\textbf{Influence of slip and viscous effects:} As shown in Fig.~\ref{fig:mesh_study_velocity_fs3d_staggered_slip_study_alpha1}, where the staggered slip length $\tilde{L}$ for $\omega=2/3$ is varied from $100\, \text{nm}$ to $500 \, \text{nm}$, the choice of the (staggered) slip length shows only a minor influence on the breakup dynamics. This behavior is to be expected in the inviscid regime (low $\Oh$, \changeC{here $\Oh = 4.7 \cdot 10^{-3}$}), where viscous dissipation due to slip is irrelevant for the breakup process. A slight monotonic decrease of the breakup speed with the slip length is visible when the breakup speed approaches the local minimum at approximately $45 \mum$. In all subsequent numerical simulations, we apply a staggered slip length of $500 \, \text{nm}$ and $\omega=2/3$. In Appendix~\ref{section:staggered_slip_appendix} it is shown that the choice of $\omega$ has no relevant influence on the breakup dynamics. The influence of the dynamic viscosity on the dynamics in the simulation is studied in Fig.~\ref{fig:alpha_1_viscosity_study}. The green curve is computed with an $50\%$ increased dynamic viscosity in both the liquid and the gas phase. Despite this large increase in the viscosity, there is only a small decrease in the breakup speed of approximately $4\%$. In particular, the overall shape is unchanged including the position of the local minimum and the dynamic exponents. From this, we conclude that viscous effects including boundary slip (see Fig.~\ref{fig:mesh_study_velocity_fs3d_staggered_slip_study_alpha1}) play no significant role \changeD{in the considered range of parameters}. Instead, the process is controlled mainly by a balance of inertial and capillary forces (see also Fig.~\ref{fig:alpha_1_breakup_velocity_rho_study_summary}).\\

\textbf{Influence of the wetting conditions:} The influence of the wetting conditions through the contact angle $\thetaphob \in \{ 90^\circ, 102^\circ, 110^\circ \} $ on the hydrophobic stripe is studied in Fig.~\ref{fig:theta_study_alpha_1}. Note that we employed Surface Evolver to compute individual initial liquid surfaces for each case. The results show that the dynamics stays qualitatively similar, while the whole graph is shifted to the right with decreasing $\thetaphob$.
The colored lines show fits to the second dynamic regime. Note that the difference in the exponents is very small. \changeD{The numerical values for the exponent $\nu$ and the constant $c$ obtained from the phase space analysis are given by 
\begin{align*}
\thetaphob = 90^\circ: \nu = 0.47596, \quad c &= 179.3147 \mum \\
\thetaphob = 102^\circ: \nu = 0.47553, \quad c &= 171.9538 \mum \\
\thetaphob = 110^\circ: \nu = 0.47623, \quad c &= 162.5795 \mum.
\end{align*}
These values have been obtained from a least-squares fit of the data starting from the local minimum in the breakup speed. Note that the last data point in Fig.~\ref{fig:theta_study_alpha_1} has been excluded from the fit for $\thetaphob=102^\circ$ and $\thetaphob=110^\circ$. Similarly, the last two data points have been excluded from the fit for $\thetaphob = 90^\circ$. In summary,} it can be concluded that the wetting condition, i.e. the contact angle \changeD{on the hydrophobic stripe}, has no major influence on the breakup dynamics in the second dynamic regime.

\subsubsection{The case $\alpha=0.5$} 
\begin{figure}[ht]
\subfigure{\includegraphics[width=8cm]{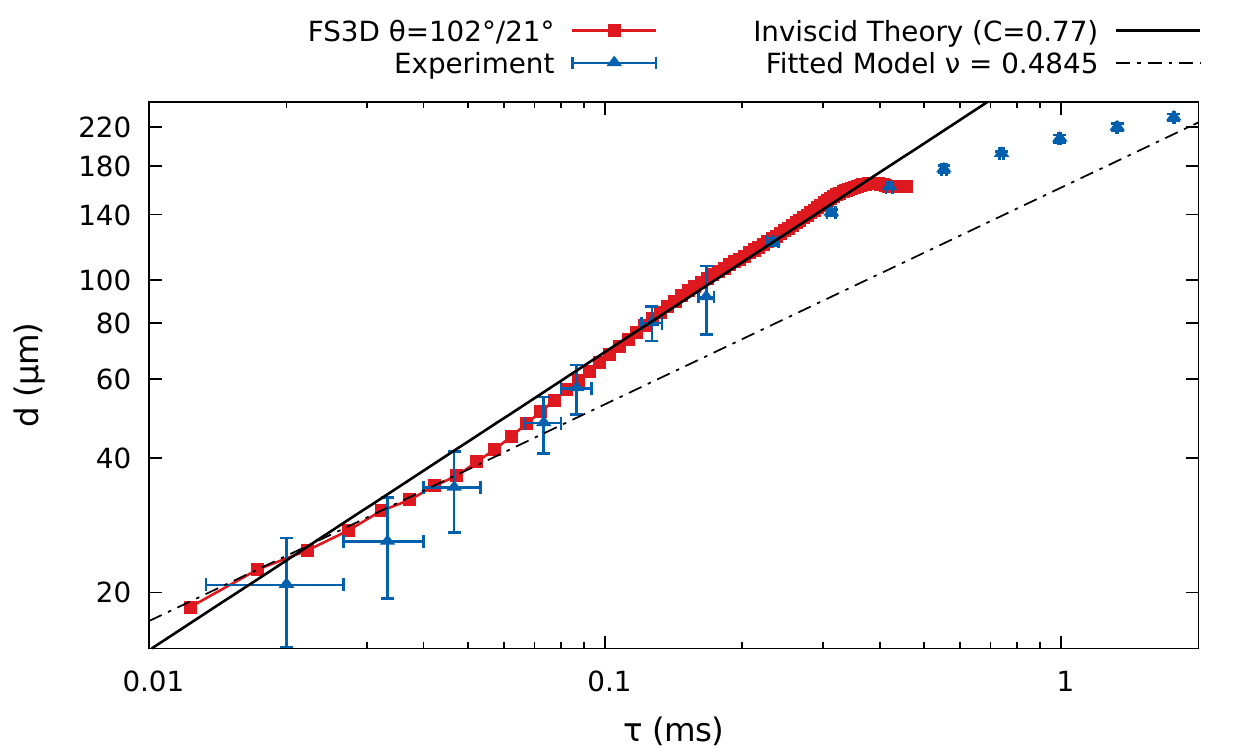}}
\subfigure{\includegraphics[width=8cm]{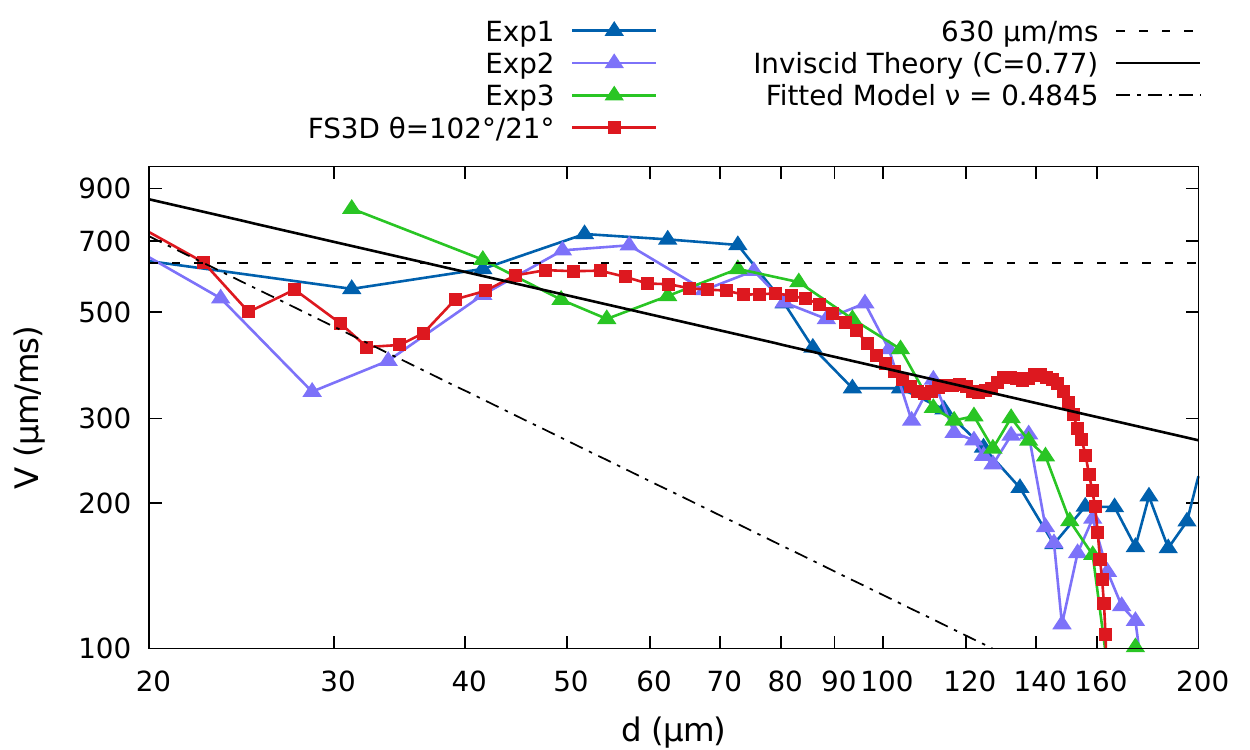}}
\caption{Breakup dynamics for $\alpha = 0.5$. Left: Minimum bridge width as a function of time. Right: Breakup speed as a function of the bridge width.}
\label{fig:mesh_study_fs3d_staggered_slip_study_alpha0.5}
\end{figure}

The breakup dynamics for $\alpha = 0.5$ is shown in Fig.~\ref{fig:mesh_study_fs3d_staggered_slip_study_alpha0.5}. The overall dynamics is qualitatively similar to the case $\alpha=1$. After a short initial phase, a good quantitative agreement between the experimental data and the numerical data is found. In the region $90 \mum \lesssim d \lesssim 160 \mum$, the dynamics is consistent to the prediction of the inviscid theory with $C \approx 0.77$ in an averaged sense. At a bridge width of about $60\mum$, the experimental breakup speed reaches a maximum at approximately $630 \mum/\text{ms}$, which is captured accurately by the numerics. For smaller bridge widths the velocity decreases until a local minimum is reached. In the numerical data, the local minimum lies at approximately 32~$\mum$, while the minimum in the experiment is shifted to smaller widths (except for Exp3). After the minimum is reached, a second dynamic regime is initiated which is similar to the one obtained for $\alpha = 1$. \changeD{A least-squares fit of the numerical data in the phase space representation yields the exponent $\nu \approx 0.48$ for the second dynamic regime.}\\

\subsubsection{The case $\alpha=1.5$} 
\begin{figure}[ht]
\subfigure{\includegraphics[width=8cm]{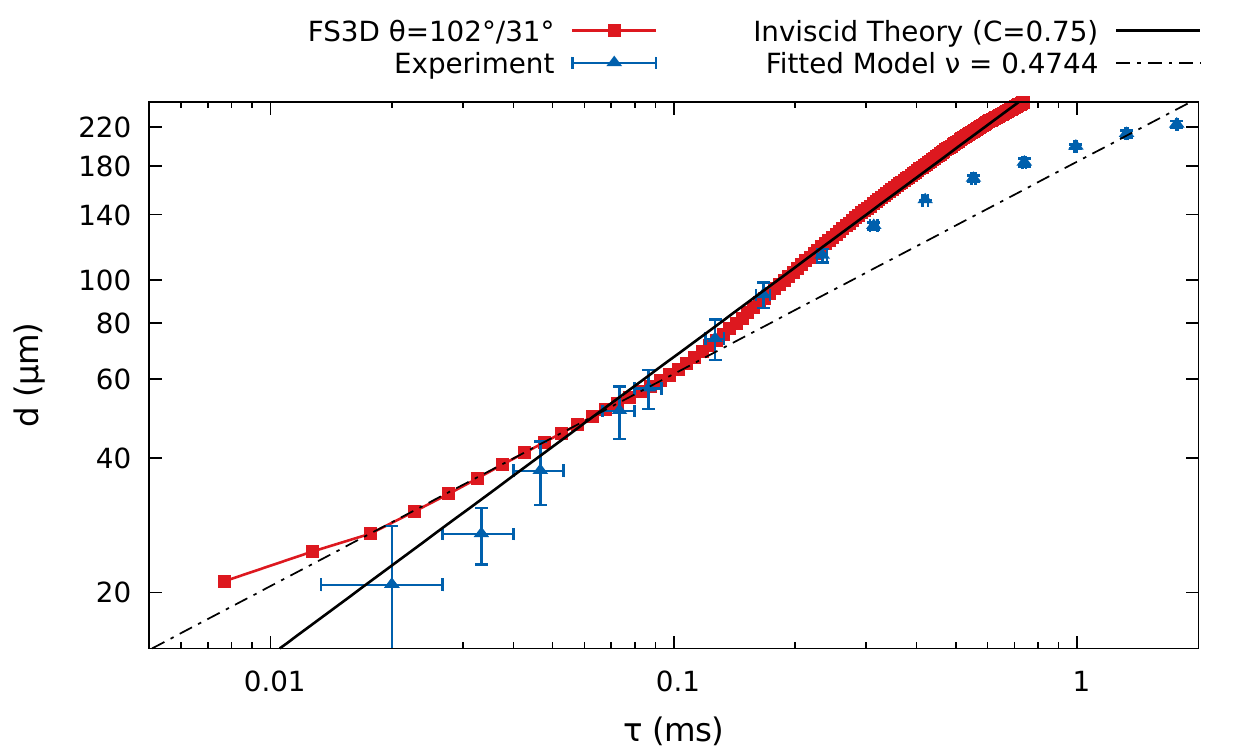}}
\subfigure{\includegraphics[width=8cm]{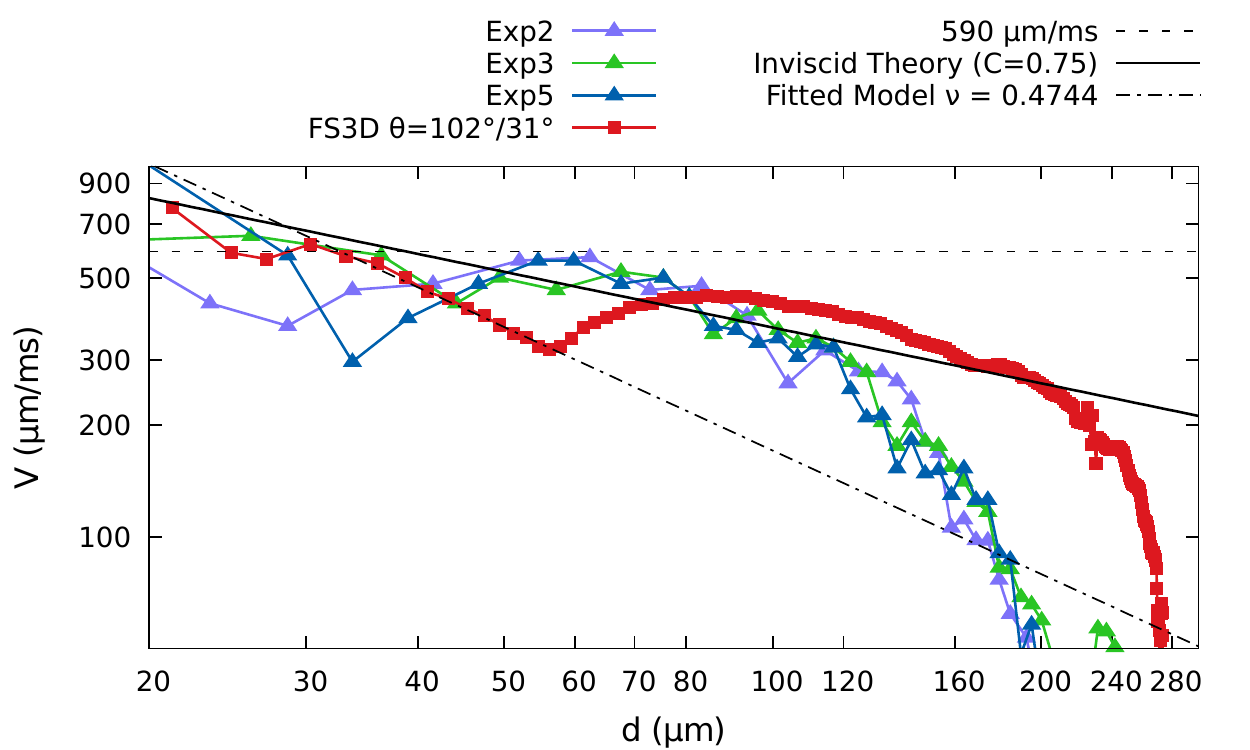}}
\caption{Breakup dynamics for $\alpha = 1.5$. Left: Minimum bridge width as a function of time. Right: Breakup speed as a function of the bridge width.}
\label{fig:mesh_study_fs3d_staggered_slip_study_alpha1.5}
\end{figure}

The results for $\alpha = 1.5$ are displayed in Fig.~\ref{fig:mesh_study_fs3d_staggered_slip_study_alpha1.5}. The dynamics of the breakup process in the simulation is qualitatively similar to the cases $\alpha = 1$ and $\alpha = 0.5$. Shortly after the onset of the instability at approximately $280 \mum$, the dynamics in the numerics is (on average) consistent with the inviscid theory for $C \approx 0.75$. At approximately $90 \mum$ the breakup speed reaches a maximum before it decreases toward a local minimum at approximately $56 \mum$. 
Below this length scale, a second dynamic regime occurs. \changeD{The observed exponent $\nu \approx 0.47$ is very close to the values observed for $\alpha=1$ and $\alpha=0.5$. Interestingly, the numerical data for the breakup speed show another local maximum at $d \approx 30 \mum$.}\\
\\
In contrast to the cases $\alpha = 1$ and $\alpha = 0.5$, there is a significant discrepancy between experiment and simulation at the onset of the breakup process. While the breakup process in the simulation starts at approximately $280 \mum$, the onset of instability in the experiments occurs at a width of the capillary bridge of approximately $200 \mum$. It appears that the whole process is shifted towards smaller values of the bridge width compared to the numerical data. \changeC{In particular, in the simulation the position of the local minimum is significantly shifted towards larger values of the minimum width.} In fact, the initial shape of the droplet delivered by Surface Evolver shows a significant deviation from the experimentally observed shape. Notably, there is a different curvature of the interface in the plane \changeC{of} the substrate \changeC{for all investigated values of $\alpha$}; see Fig.\ref{fig:initCondition} in Appendix~\ref{section:initial_condition}.
\changeC{In the experiments, the volume decreases due to evaporation before the critical bridge width is reached.
Consequently, the minimum width decreases and the pressure within the capillary bridge increases.
Shortly before the point of instability is reached, the capillary system responds to the increased pressure within the bridge by increasing the curvature, and therefore the pressure, above the hydrophilic stripes.
As a consequence, the contact angle $\Theta_\mathrm{phil}$ increases, since liquid is pumped from the region of highest pressure (location of minimum bridge width) towards the liquid reservoirs above the hydrophilic stripes (see~\cite{Hartmann2019}).
This process cannot be captured by the Surface Evolver simulations and, as a consequence, the initial curvature in the numerical simulations deviates from the one in the experiments when the capillary structure reaches the point of instability.
} 
This deviation might explain the shift in the dynamics. 
\subsection{Rayleigh-Plateau Instability}
\label{sec:rayleigh_plateau}
\begin{figure}[hbt]
 \centering
 \includegraphics[width=16.5cm]{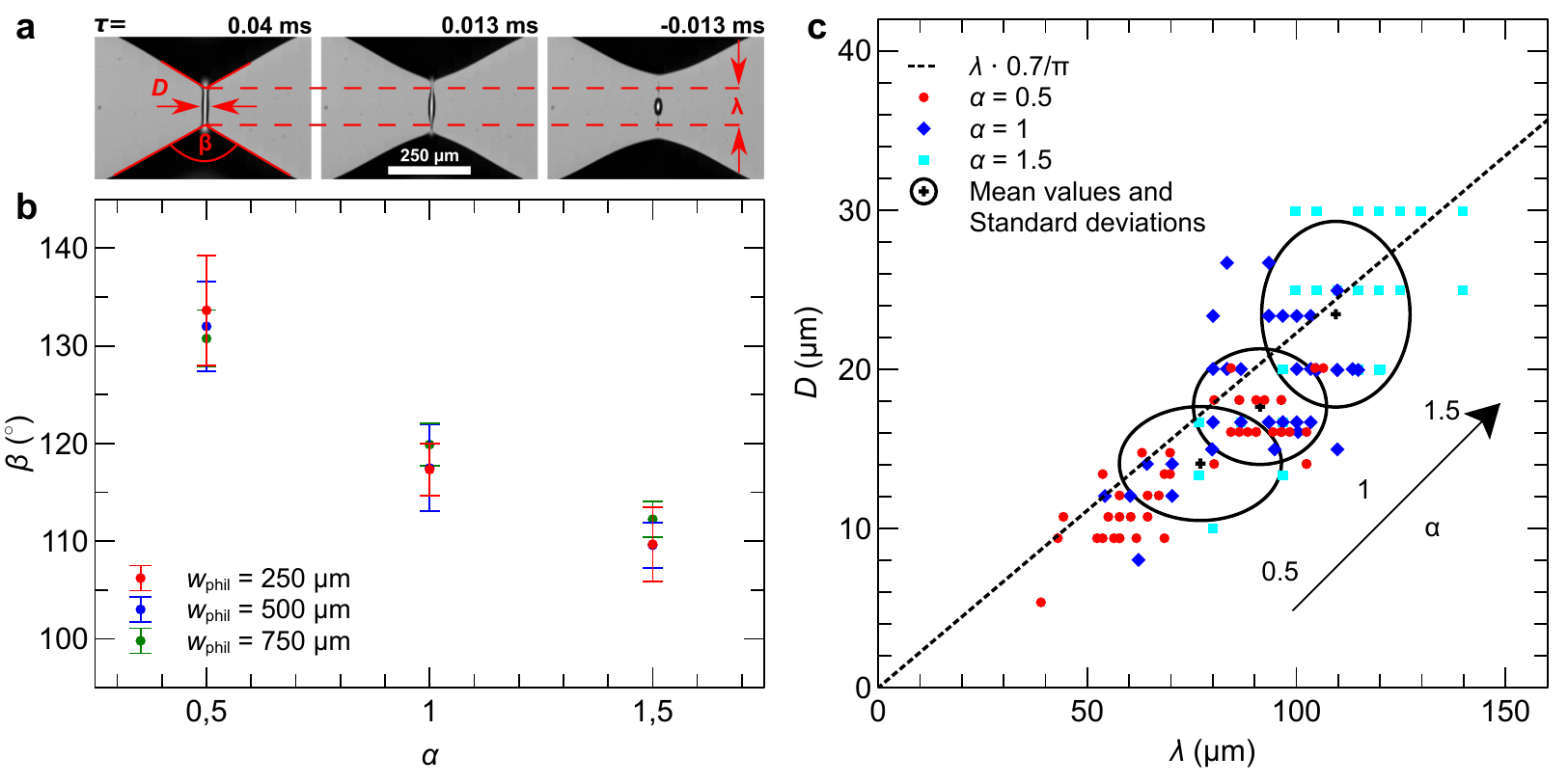}
 \caption{Analysis of the final breakup behavior. In (a), the capillary bridge has evolved into a liquid cylinder between two cones. The final breakup happens at two pinch-off points leaving behind a primary droplet with smaller secondary droplets in the middle of the hydrophobic stripe. In (b), the cone's opening angles $\beta$ are evaluated for different $\alpha$'s and hydrophilic stripe widths. In (c), the diameter $D$ of the cylindrical segment is plotted versus the measured wavelength $\lambda$ for $w_\mathrm{phil} = 500 \mum$ and different $\alpha$. Each point represents one experiment. These are condensed in the ellipses, representing the standard deviation around the mean value for each $\alpha$. The dashed line corresponds to \changeC{equation}~\eqref{eq:rayleighPlateau}.}
 \label{fig:rayleighPlateau_combi}
\end{figure}

In Fig.~\ref{fig:rayleighPlateau_combi}a, the final moments of bridge breakup are shown for $w_\mathrm{phil}=500 \mum$ and $\alpha = 1$.
Emerging from a catenoid type structure, as already shown in Fig.~\ref{fig:qualitativeComparison}, a liquid thread (a cylinder cut by a plane) with diameter $D$ is formed between the tips of two cones with opening angle $\beta$ ($\tau = 0.04$~ms).
The thread then gets pinched at the tips of the cones ($\tau = 0.013$~ms) and finally breaks up, leaving a primary droplet in the middle of the hydrophobic stripe ($\tau = -0.013$~ms).
Besides the primary droplet, some smaller secondary droplets can be seen in the experiments.
These droplets have their origin in a self-similar breakup process.
This phenomenon is also found in the breakup of a soap film that is spanned between two circular rings~\cite{Chen1997}, \changeC{or in the breakup of a liquid jet~\cite{Kowalewski1996,Eggers1997}.}\\

The described breakup process exhibits two correlated features:
First of all, cones, which in the two-dimensional projection appear as triangles, are formed in each experiment (at all observed values of $\alpha$) at a certain point in time.
This is close to the point where a liquid thread is formed.
As can be seen in Fig.~\ref{fig:rayleighPlateau_combi}b, these cones have unique opening angles~$\beta$ for each $\alpha$. 
Their structure is independent of the hydrophilic stripe width.
The value of $\beta$ is highly reproducible between different experiments.
\\
The liquid thread breaks up in a self-similar process.
Fig.~\ref{fig:rayleighPlateau_combi}c shows all data of experiments performed with a hydrophilic stripe width of $500~\mum$ for $\alpha = 0.5 - 1.5$.
The diameter $D$ of the cylindrical thread is plotted versus the parameter $\lambda$, which is the distance between the two pinch-off points and simultaneously the distance between the two cone tips in Fig.~\ref{fig:rayleighPlateau_combi}a.
Note that these data points, as well as the ones depicted in figure~\ref{fig:rayleighPlateau_combi}b, originate from different experiments that were performed using a microscope.
This setup has the advantage of a higher optical resolution.
As a drawback, it does not allow side view imaging of the droplet.
However, the side view data is needed for the comparison of the experiments with the simulations.
Different colors are used for different $\alpha$'s, and the mean values and standard deviations for each $\alpha$ are represented by ellipses around black crosses.
The values are compared with the classical Rayleigh-Plateau instability, which predicts that a cylindrical liquid jet of diameter $D$ becomes unstable since a decomposition into droplets is energetically more favorable.
$D$ is connected to the wavelength $\lambda$ that belongs to the \changeC{fastest growing} mode according to the formula \changeC{(see e.g.~\cite{Drazin2002,Eggers2008,Eggers2015})}
\begin{equation}
\changeC{\frac{\pi D}{\lambda} \approx 0.7}.
\label{eq:rayleighPlateau}
\end{equation}
From Fig.~\ref{fig:rayleighPlateau_combi}c it can be seen that, within the experimental error, the final breakup process on the substrate follows the classical Rayleigh-Plateau instability.
The measured receding contact angle is 102$^\circ$, which is slightly higher than 90$^\circ$.
For the case of 90$^\circ$ a liquid jet with unpinned contact lines becomes unstable following the classical Rayleigh-Plateau result (see, e.g., Bostwick and Steen~\changeC{\cite{Bostwick2015,Bostwick2018}}).
This also follows from symmetry considerations.
Remarkably, the wavelength \changeC{of the fastest growing mode} corresponds to the distance between the tips of the cones and the distance between the secondary satellite droplets.
The reasonable agreement with the classical Rayleigh-Plateau theory, which is based on an inviscid fluid, also indicates that the influence of viscous stresses that shows up in the final stages of breakup, is not significant for the decay of the thread.
The scattering of the data points of each experiment might be due to the uncertainty in the receding contact angle.
This is already discussed in Section~\ref{sec:qualitativeComparison}.
Furthermore, a part of the data scatter visible in Fig.~\ref{fig:rayleighPlateau_combi}c is due to the spatial resolution of the camera.
Since the final breakup process occurs at a relatively small time scale ($\approx$0.04~ms in case of $w_\mathrm{phil}$=500~$\mu$m and $\alpha$ = 1, see Fig.~\ref{fig:rayleighPlateau_combi}a), a high frame rate is needed for image acquisition.
Additionally, the region of interest is smaller than 1~mm. In order to achieve a good image quality with the given light source, image acquisition is limited to the given pixel resolution.
 
\section{Conclusion}
The breakup dynamics of a capillary bridge on a hydrophobic stripe, which forms during evaporation of a droplet, was studied experimentally and numerically.
The droplet wets two hydrophilic stripes, separated by a hydrophobic stripe.
Different ratios of the hydrophobic and hydrophilic stripe width $\alpha$ were considered.
By performing experiments with two synchronized high-speed cameras for $\alpha = $ 0.5, 1 and 1.5, the breakup dynamics of the capillary bridge could be observed simultaneously to the contact angle change on the hydrophilic stripe.
Consequently, the contact angle data could be used to calculate physically realistic initial conditions for the continuum mechanical simulations using Surface Evolver.
\\
\\
The data import into the numerical simulation tool Free Surface 3D (FS3D) was achieved using an algorithm within OpenFOAM that converts the surface mesh from Surface Evolver into a volume fraction field.
The geometric Volume-of-Fluid method was employed to solve the three-dimensional two-phase Navier Stokes equations.
In order to dampen spurious currents at the contact line, the Boundary Youngs interface reconstruction algorithm~\cite{Fricke.2019b} was adapted to three dimensions and a modification of the Navier Slip boundary condition was introduced (``staggered slip'').\\
\\
The phase space picture of the breakup dynamics, that allows for a detailed \changeC{and systematic} study of the \changeC{dynamics} of the process, is employed.
In particular, the \changeC{ambiguity related to the} choice of the breakup time $t_0$ is removed. Two distinct regimes were observed both in the experiments and the numerical simulations which are connected by a transition region. The initial regime follows approximately \changeD{(in an average sense)} the well-known inviscid relation for a \emph{free} capillary bridge, i.e. $d(\tau)  \propto C (\sigma \tau^2 / \rho)^{1/3}$, where $C$ is found to be $0.75 - 0.77$ in both experiments and simulations. The regime transition is characterized by a maximum in the breakup speed followed by a local minimum which corresponds to the onset of the second dynamic regime.
Within the second regime, a different exponent $\nu$ can be determined ranging from  \changeD{$0.47$ to $0.48$} in the investigated range of scales and parameters. The latter values differ significantly from those obtained from dimensional analysis for a free capillary bridge in the inviscid regime. This means that either the presence of the moving contact line affects the exponent or that the final universal regime is not yet reached. \changeD{Nevertheless, it is shown that the complex dynamics is governed by a balance of inertial and capillary forces. Viscous forces and boundary slip show only a minor influence on the time evolution of the minimum width.} Experiments and simulations show a good quantitative agreement in terms of the breakup speed down to a bridge width of approximately $20 \mum$.\\
\\
In both experiments and simulations, a liquid thread (a cylinder cut by a plane) is formed between two cones.
These have a well-defined opening angle that decreases with increasing $\alpha$.
From the experiments, it could be inferred that the final liquid thread breaks up in a Rayleigh-Plateau type instability.
Remarkably, the wavelength of the breakup \changeC{belonging to the fastest growing mode} corresponds to the distance between the tips of the cones.

\paragraph{Acknowledgments:} We kindly acknowledge the financial support by the German Research Foundation (DFG) within the Collaborative Research Centre 1194 “Interaction of Transport and Wetting Processes” – Project-ID 265191195, subprojects A02b and B01, B02 and Z-INF. Calculations for this research were conducted on the Lichtenberg high performance computer of the TU Darmstadt.\\
\\
Maximilian Hartmann and Mathis Fricke contributed equally to this work. 

\appendix
\section{On the Staggered Slip Boundary Condition}
\label{section:staggered_slip_appendix}
\changeB{The staggered slip condition introduced in Section~\ref{subsection:vof} reduces the amount of artificial numerical slip and, thereby, is able to reduce spurious velocities at the domain boundary and at the contact line. The method is conceptually simple, can be implemented in a single line of code and is surprisingly effective (see below). The purpose of this section is to give some more details on the approach and its effect on dynamic wetting simulations. A more comprehensive discussion can be found in Chapter~11 of \cite{Fricke2021}.}\\
\\
\changeB{The motivation for the adaptation of the Navier slip condition is the observation that the contact line speed is usually over-predicted on coarse meshes. There are (at least) two reasons for this numerical phenomenon: The first reason is that the interface in a cell located at the boundary is advected by a face-centered velocity located at a distance $\Delta x/2$ above the physical boundary. The latter velocity is typically much larger than the expected velocity right at the boundary leading to an over-prediction of the contact line speed. The second reason is that the loss of energy due to dissipation in the region close to the contact line is not properly resolved on coarse meshes. This ``missing dissipation'' is mimicked by the increased discrete viscous dissipation in a boundary cell caused by the increased opposite velocity in the ghost cell resulting from the staggered slip condition. Clearly, the proposed method is somewhat ad hoc and the results could probably be further improved by a more sophisticated numerical modeling of the missing viscous dissipation. Nevertheless, the method is able to improve the convergence in a dynamic wetting simulation on coarse meshes when the slip length is not resolved by the computational mesh (see below). }

\begin{figure}[hb]
\centering
\subfigure[$\omega=0$]{\includegraphics[width=0.42\textwidth]{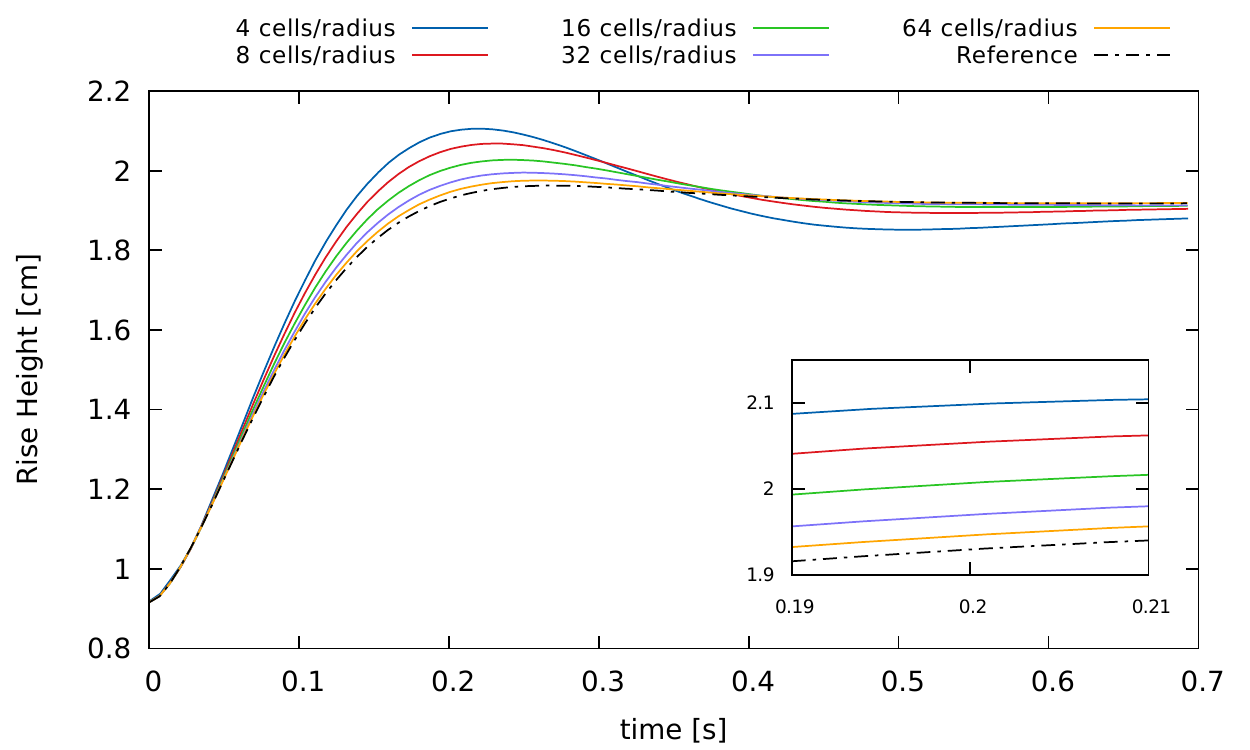}}
\subfigure[$\omega=1$]{\includegraphics[width=0.42\textwidth]{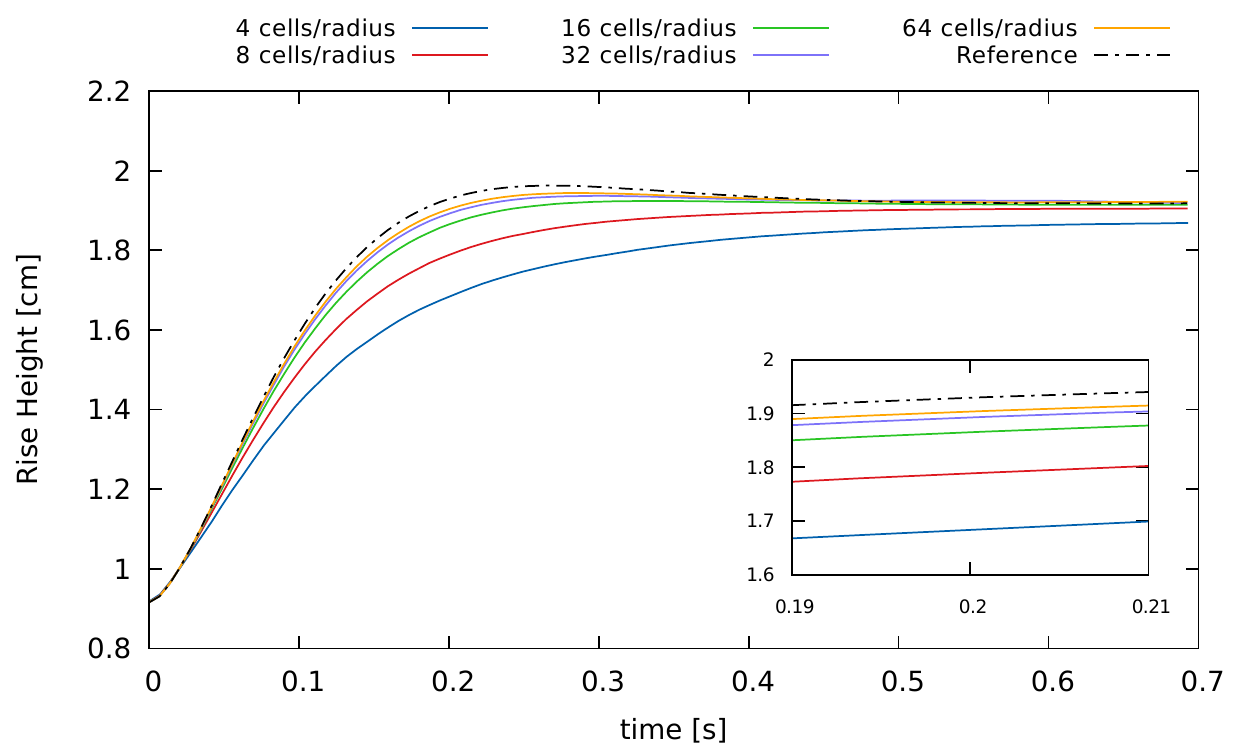}}
\subfigure[$\omega=1/2$]{\includegraphics[width=0.42\textwidth]{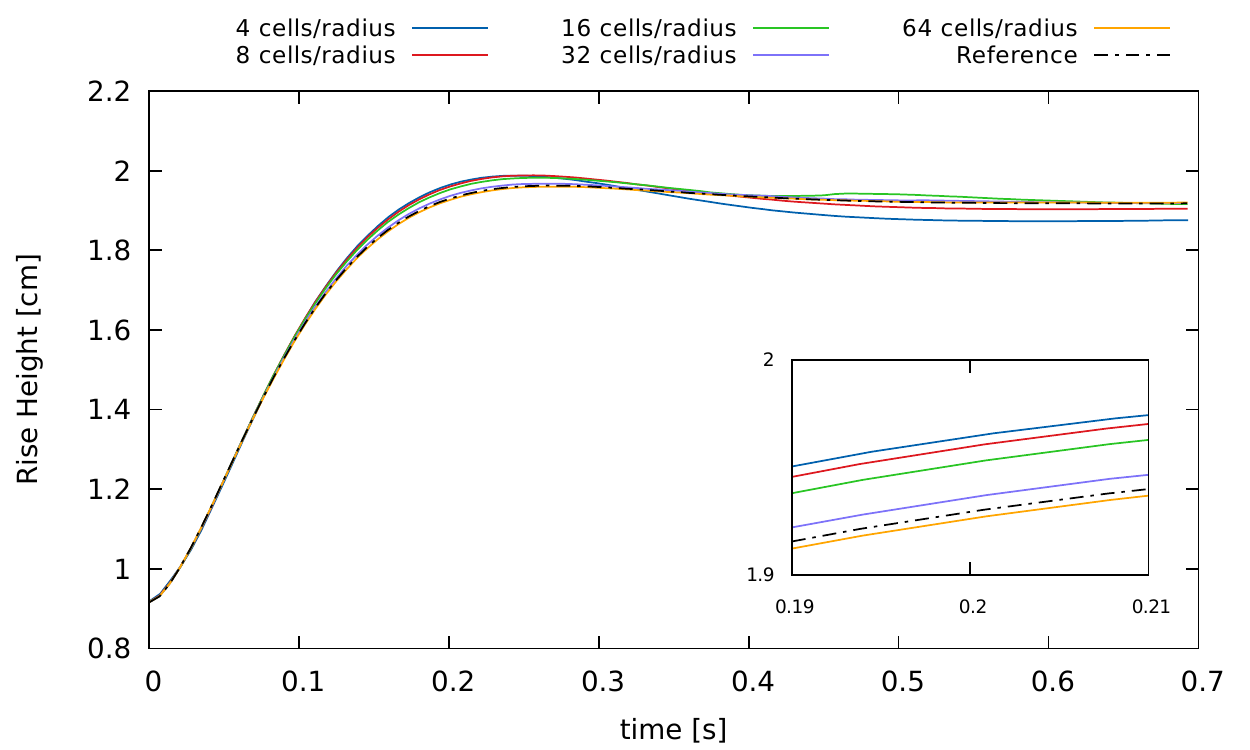}}
\subfigure[Convergence in the maximum norm.]{\includegraphics[width=0.42\textwidth]{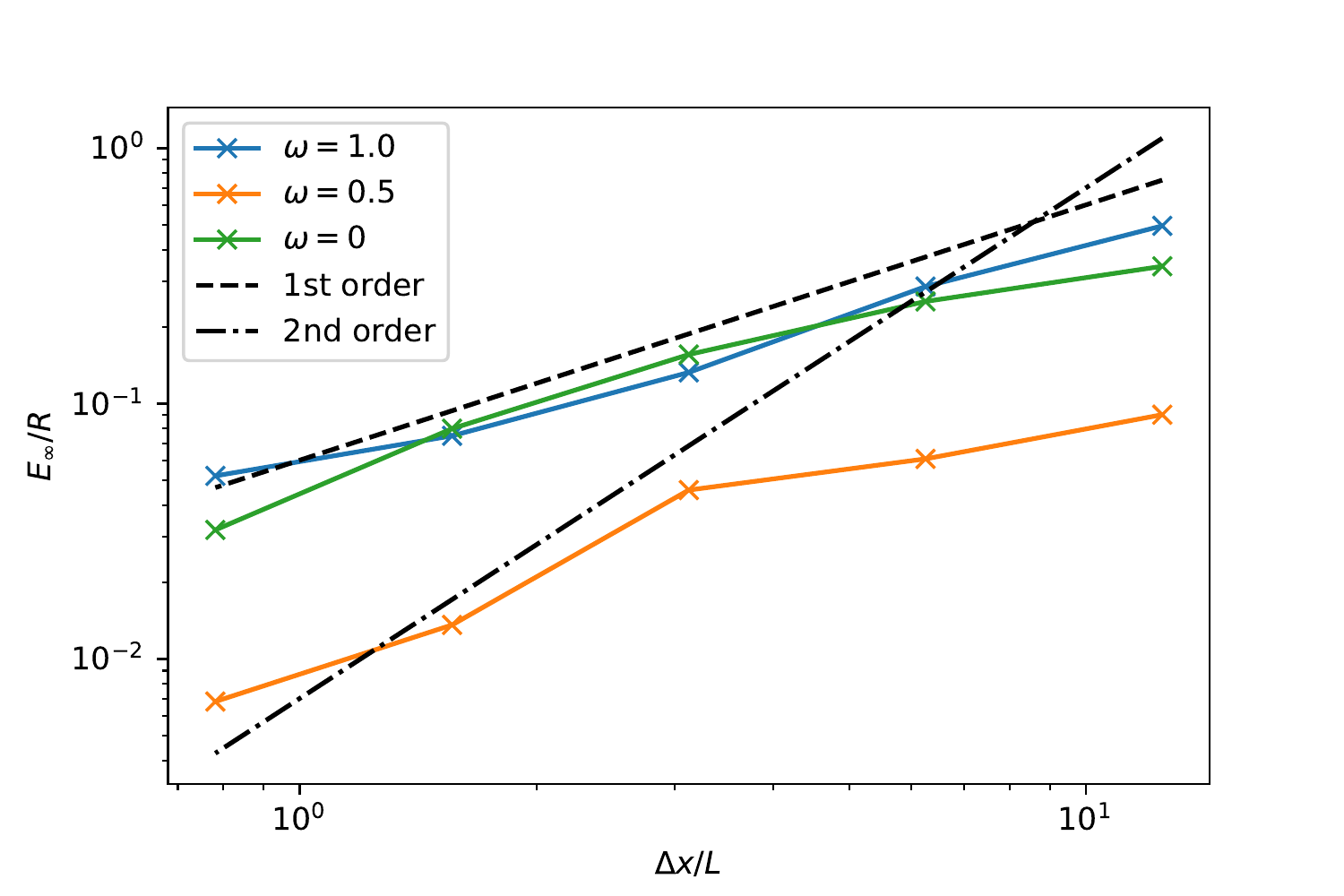}\label{fig:capillary_rise/convergence}}
\caption{\changeB{Capillary rise dynamics using the standard ($\omega=0$) and the staggered ($\omega=1/2,1$) slip boundary condition compared to reference data by Gründing et al.\ \cite{Gruending.2020,Gruending.2020-data} for $L=R/50$ and $\Omega=1$}.}
	\label{fig:capillary_rise}
\end{figure} 

\paragraph{Capillary rise benchmark:} We examine the effect of the staggered slip boundary condition for the numerical simulation of wetting flows based on the capillary rise problem as described in detail by Gründing et al.\ \cite{Gruending.2020,Gruending.2020-data}. The latter study provides a comprehensive numerical dataset on the dynamics of the two-dimensional capillary rise in different regimes ranging from a highly oscillatory behavior to a purely monotonic rise towards the stationary state. Following the work by Fries and Dreyer \cite{Fries2009} and Gründing et al.\ \cite{Gruending.2020}, the physical parameters of the systems are varied such that different values of the non-dimensional parameter (with $R$ the radius of the capillary)
\[ \Omega = \sqrt\frac{9 \sigma \cos \theta \visc^2}{\rho^3 g^2 R^5} \]
are realized. The \changeB{numerical results for different values of $\omega$ and the} reference data \cite{Gruending.2020-data} in terms of the rise height\footnote{\changeB{The rise height is defined as the height of the meniscus relative to the liquid bath.}} over time for the setup and parameters specified by Gründing et al.\ \cite{Gruending.2020} for a slip length $L=R/50$ and $\Omega=1$ are shown in Fig.~\ref{fig:capillary_rise}. The reference data are \changeB{computed on a fine mesh resolving the slip length and are} confirmed by four different numerical methods\footnote{\changeC{The study \cite{Gruending.2020} compares the present geometrical VOF method \textit{FS3D} with the algebraic VOF method \textit{interFoam}, the Arbitrary-Lagrangian-Eulerian method \textit{interTrackFoam} and the level-set based extended discontinuous Galerkin method \textit{BoSSS}; see \cite{Gruending.2020} for more details.}} including the present implementation of FS3D with the standard Navier slip condition \eqref{eqn:ghost_velocity_navier_slip}. \changeB{For the standard Navier slip condition ($\omega=0$) it is found that the reference solution is approached from \emph{above}, i.e.\ the contact line velocity is over-estimated on coarse grids. Conversely, the reference solution is approached from \emph{below} for $\omega=1$, i.e.\ the contact line velocity is under-estimated on coarse meshes. The error in terms of the maximum deviation of the rise height over time is similar for $\omega=0$ and $\omega=1$; see Fig.~\ref{fig:capillary_rise/convergence}. The error is reduced by approximately one order of magnitude for the choice $\omega=1/2$.}

\paragraph{Single-phase channel flow:} \changeB{Another instructive example for the validation of the boundary condition is a single-phase flow in a two-dimensional channel driven by a prescribed pressure gradient $G=-\partial_x p$. An analytical solution for the Navier Stokes equations with the Navier slip boundary condition is available in this case serving as a reference for validation. In particular, one can show that the mass transport rate across the channel is proportional to $1 + 6 L/H$, where $H$ is the height of the channel. The channel flow problem is studied with the staggered slip implementation in \cite[p.~156]{Fricke2021}. It is found that the order of convergence for the mass transport rate drops from two to one when the staggered slip condition is applied (unless $\omega$ is chosen to be very small). Hence, the standard Navier slip condition is more accurate in the single-phase case; see \cite{Fricke2021} for more details.}

\begin{figure}[ht]
\centering
\includegraphics[width=0.48\textwidth]{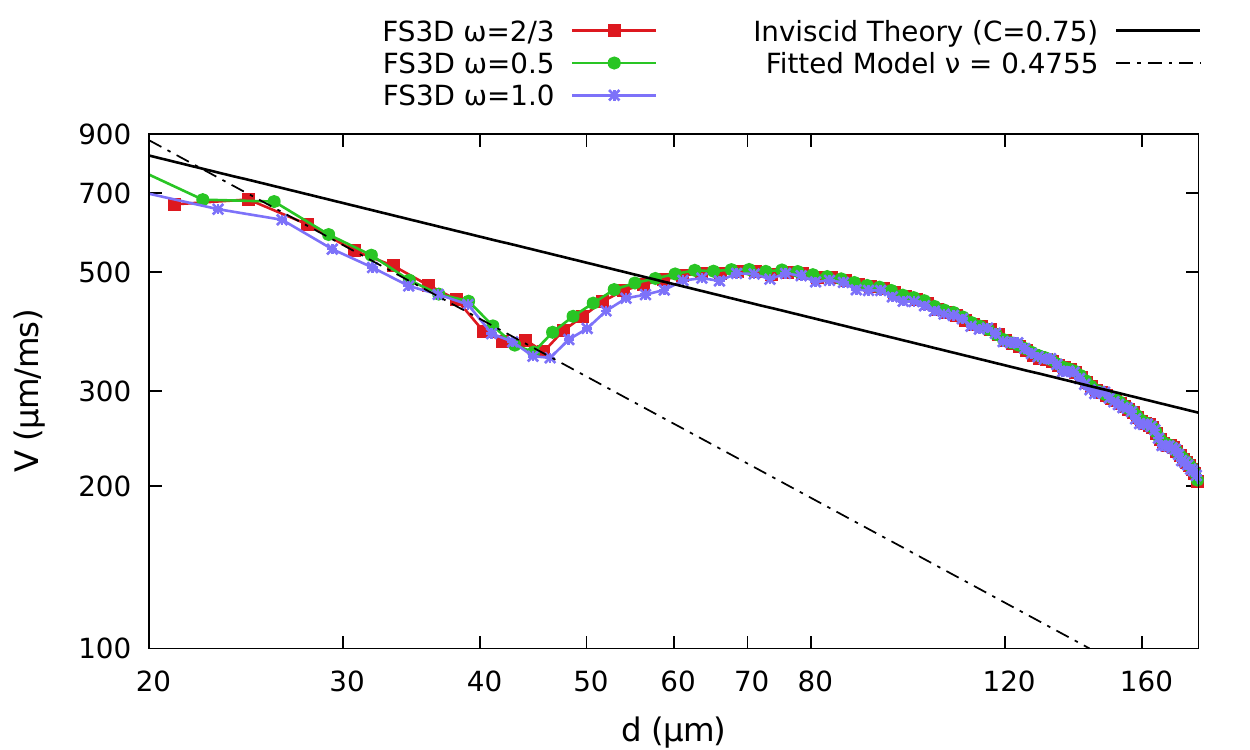} 
\caption{Influence of the choice of $\omega$ in \eqref{eqn:ghost_velocity_staggered_slip} on the breakup dynamics ($\alpha = 1$).}
\label{fig:alpha_1_staggered_slip_factor}
\end{figure}

\paragraph{Effect on the break dynamics:} \changeB{It is important to} note that in the present study of the breakup dynamics of a liquid bridge, \changeB{the staggered slip condition} only serves the purpose of damping spurious currents at the contact line. The choice of $\omega$ shows only a weak influence of the breakup dynamics in the simulation; see Fig.~\ref{fig:alpha_1_staggered_slip_factor}. The dynamic exponents remain unchanged and a difference is visible only when the local minimum in the velocity is approached. \changeB{This behavior is consistent with the observation that viscous effects show only a minor influence on the breakup process on the considered length scales.}

\changeB{In summary, the staggered slip condition is easy to implement and appears to effectively model the missing viscous dissipation in the contact line region which is not resolved by the grid. However, the optimal value for the free parameter $\omega$ cannot be estimated a priori. Further research is necessary to understand the underlying mechanisms at the moving contact line more quantitatively.}

\clearpage
 
\section{Initial Condition}
\label{section:initial_condition}
\begin{figure}[htp]
 \center
 \includegraphics[]{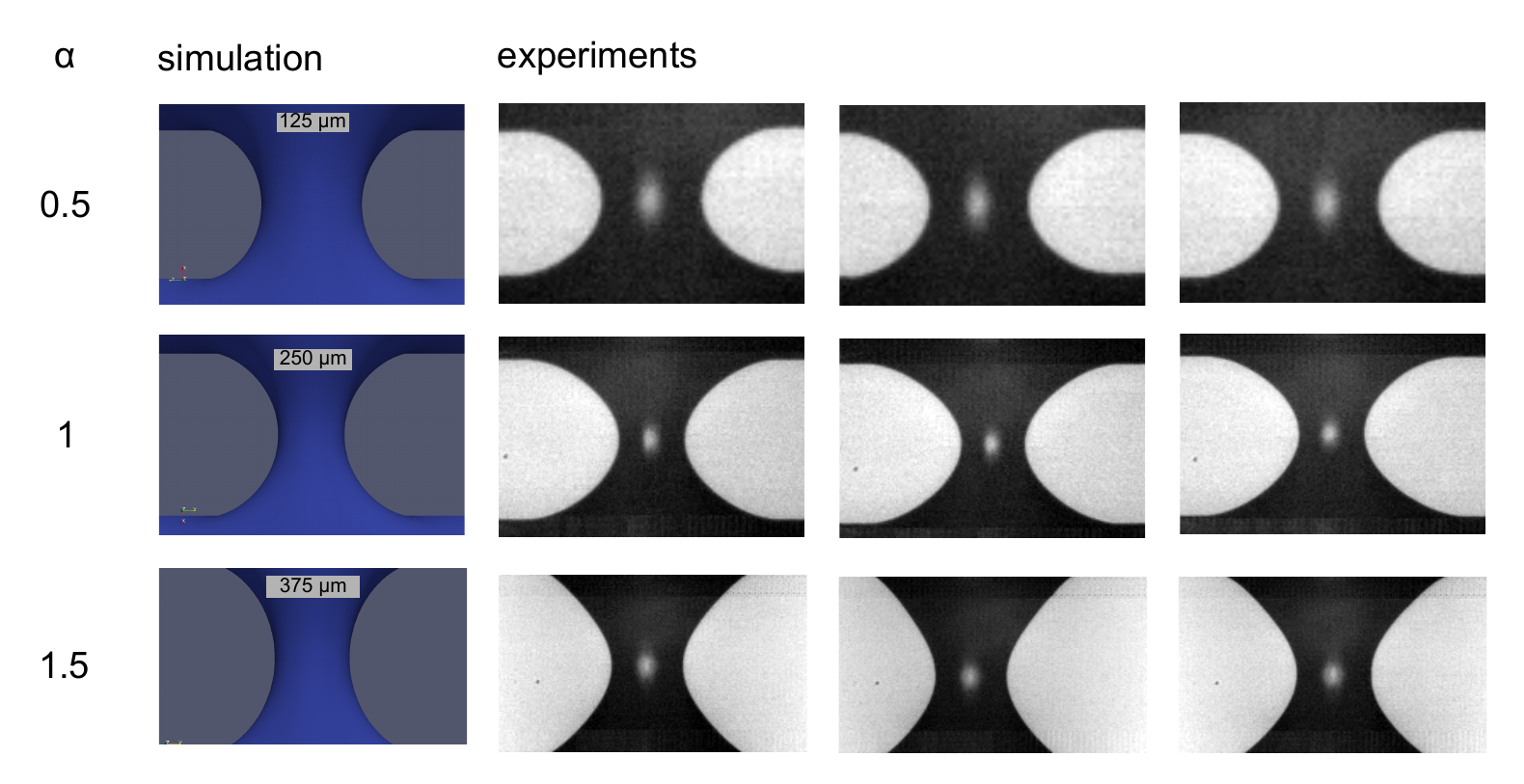}
\caption{Initial condition for the numerical simulations compared to the corresponding image from the experiments. In the first column, a scale bar is depicted, which is valid for each respective row. \changeC{In the column for the simulations, the liquid is depicted in blue, in the experimental part in black color. In both cases, the substrate is visible in gray color.}}
\label{fig:initCondition}
\end{figure}

Figure~\ref{fig:initCondition} shows the initial condition for the numerical simulations generated with Surface Evolver compared to the corresponding state of three different experiments for different $\alpha$. The experimental images shown are chosen such that the minimum width $d$ of the capillary bridge matches the initial condition.
\changeC{Obviously, the curvatures in the plane parallel to the substrate deviate between the experiments and the deviation increases for increasing $\alpha$.
This is especially obvious in the quantitative comparison between experiment and simulation for $\alpha = 1.5$.
In the experiments, the capillary surface adopts its shape shortly before the final point of instability.
However, Surface Evolver is not able to calculate this transient process, since it can only be used to calculate equilibrium shapes.
This final time span in the experiments, even though it is very short, might lead to these small deviations.
Nevertheless, employing Surface Evolver appears to be the only method to compute the initial condition with reasonable effort.}

\section{Evaporation and Capillary Time Scales}
\label{section:timescale_estimation}
We expect the breakup process to occur on the capillary time scale $T$, see equation~(\ref{eqn:inviscid_breakup_power_law}).
Without going into detail, $T$ can be estimated as $T = \sqrt{\frac{\rho}{\sigma} w_\mathrm{phob}^3}$, with $\rho$ being the density of the liquid, $\sigma$ being the surface tension and $w_\mathrm{phob}$ being the hydrophobic stripe width.
The evaporation time scale of a sessile droplet can be estimated as $T_\mathrm{evap} = \frac{\rho L H}{D_\mathrm{w,a} \Delta c_\mathrm{w}}$, with $L$ and $H$ being two different length scales, $D_\mathrm{w,a} = 2.4 \cdot 10 ^{-5}~\mathrm{m}^2/\mathrm{s}$ being the diffusion coefficient of water vapor in air and $\Delta c_\mathrm{w} \approx 1 \cdot 10^{-2}~\mathrm{kg}/\mathrm{m}^3$ being the concentration difference of a saturated atmosphere to the surrounding air~\cite{Li2018}.
For typical widths of the capillary bridge, which are in the order of 100~$\mu$m, $\frac{T_\mathrm{evap}}{T}\approx 10^{5}$ and therefore evaporation can be neglected during breakup.

\clearpage
\bibliographystyle{abbrvurl}

\end{document}